\documentclass[10pt]{article}
\usepackage{amssymb,amsmath}
\usepackage{amsthm}
\usepackage{mathrsfs}
\usepackage{dcolumn}
\usepackage{bm}
\usepackage{fullpage}
\usepackage{color}
\usepackage[all]{xy}
\usepackage[pdftex]{graphicx,hyperref}
\usepackage{ulem}
\usepackage{ascmac}
\usepackage{caption}
\usepackage{here}
\usepackage{comment}
\usepackage{tikz}

\DeclareMathOperator{\dr}{d}

\DeclareMathOperator{\bdr}{\mathrm{\mathbf d} }
\DeclareMathOperator{\sstar}{\divideontimes}

\begin{document}
\theoremstyle{definition}
\newtheorem{Definition}{Definition}[section]
\newtheorem{Theorem}[Definition]{Theorem}
\newtheorem*{theorem}{Theorem}
\newtheorem{Proposition}[Definition]{Proposition}
\newtheorem{Question}[Definition]{Question}
\newtheorem{Lemma}[Definition]{Lemma}
\newtheorem*{Proof}{Proof}
\newtheorem{Example}[Definition]{Example}
\newtheorem{Postulate}[Definition]{Postulate}
\newtheorem{Corollary}[Definition]{Corollary}
\newtheorem{Remark}[Definition]{Remark}
\newtheorem{Claim}[Definition]{Claim}
\newtheorem{Assumption}[Definition]{Assumption}

\theoremstyle{remark}
\newcommand{\beq}{\begin{equation}}
\newcommand{\beqa}{\begin{eqnarray}}
\newcommand{\eeq}{\end{equation}}
\newcommand{\eeqa}{\end{eqnarray}}
\newcommand{\non}{\nonumber}
\newcommand{\lb}{\label}
\newcommand{\fr}[1]{(\ref{#1})}
\newcommand{\bb}{\mbox{\boldmath {$b$}}}
\newcommand{\bbe}{\mbox{\boldmath {$e$}}}
\newcommand{\bt}{\mbox{\boldmath {$t$}}}
\newcommand{\bn}{\mbox{\boldmath {$n$}}}
\newcommand{\br}{\mbox{\boldmath {$r$}}}
\newcommand{\bC}{\mbox{\boldmath {$C$}}}
\newcommand{\bp}{\mbox{\boldmath {$p$}}}
\newcommand{\bx}{\mbox{\boldmath {$x$}}}
\newcommand{\bF}{\mbox{\boldmath {$F$}}}
\newcommand{\bT}{\mbox{\boldmath {$T$}}}
\newcommand{\bQ}{\mbox{\boldmath {$Q$}}}
\newcommand{\bS}{\mbox{\boldmath {$S$}}}
\newcommand{\balpha}{\mbox{\boldmath {$\alpha$}}}
\newcommand{\bomega}{\mbox{\boldmath {$\omega$}}}
\newcommand{\ve}{{\varepsilon}}
\newcommand{\e}{\mathrm{e}}
\newcommand{\f}{\mathrm{f}}
\newcommand{\m}{\mathrm{m}}
\newcommand{\s}{\mathrm{s}}
\newcommand{\B}{\mathrm{B}}
\newcommand{\E}{\mathrm{E}}
\newcommand{\G}{\mathrm{G}}
\newcommand{\R}{\mathrm{R}}
\newcommand{\Z}{\mathrm{Z}}
\newcommand{\HH}{\mathrm{H}}
\newcommand{\I}{\mathrm{I}}
\newcommand{\II}{\mathrm{II}}
\newcommand{\Id}{\mathrm{Id}}
\newcommand{\ddiv}{\mathrm{div}}
\newcommand{\Ising}{\footnotesize\mathrm{Ising}}
\newcommand{\std}{\mathrm{std}}
\newcommand{\hF}{\widehat F}
\newcommand{\hL}{\widehat L}
\newcommand{\tA}{\widetilde A}
\newcommand{\tB}{\widetilde B}
\newcommand{\tC}{\widetilde C}
\newcommand{\tL}{\widetilde L}
\newcommand{\tK}{\widetilde K}
\newcommand{\tX}{\widetilde X}
\newcommand{\tY}{\widetilde Y}
\newcommand{\tU}{\widetilde U}
\newcommand{\tZ}{\widetilde Z}
\newcommand{\talpha}{\widetilde \alpha}
\newcommand{\te}{\widetilde e}
\newcommand{\tv}{\widetilde v}
\newcommand{\ts}{\widetilde s}
\newcommand{\tx}{\widetilde x}
\newcommand{\ty}{\widetilde y}
\newcommand{\ud}{\underline{\delta}}
\newcommand{\uD}{\underline{\Delta}}
\newcommand{\chN}{\check{N}}
\newcommand{\cA}{{\cal A}}
\newcommand{\cB}{{\cal B}}
\newcommand{\cC}{{\cal C}}
\newcommand{\cD}{{\cal D}}
\newcommand{\cE}{{\cal E}}
\newcommand{\cF}{{\cal F}}
\newcommand{\cG}{{\cal G}}
\newcommand{\cH}{{\cal H}}
\newcommand{\cI}{{\cal I}}
\newcommand{\cJ}{{\cal J}}
\newcommand{\cK}{{\cal K}}
\newcommand{\cL}{{\cal L}}
\newcommand{\cM}{{\cal M}}
\newcommand{\cN}{{\cal N}}
\newcommand{\cO}{{\cal O}}
\newcommand{\cP}{{\cal P}}
\newcommand{\cQ}{{\cal Q}}
\newcommand{\cS}{{\cal S}}
\newcommand{\cT}{{\cal T}}
\newcommand{\cY}{{\cal Y}}
\newcommand{\cU}{{\cal U}}
\newcommand{\cV}{{\cal V}}
\newcommand{\cW}{{\cal W}}
\newcommand{\vecX}{\mathfrak{X}}
\newcommand{\tcA}{\widetilde{\cal A}}
\newcommand{\DD}{{\cal D}}
\newcommand\TYPE[3]{ \underset {(#1)}{\overset{{#3}}{#2}}  }
\newcommand{\Qc}{\overset{\footnotesize\circ}{Q}}
\newcommand{\bfe}{\boldsymbol e} 
\newcommand{\bfb}{{\boldsymbol b}}
\newcommand{\bfd}{{\boldsymbol d}}
\newcommand{\bfh}{{\boldsymbol h}}
\newcommand{\bfj}{{\boldsymbol j}}
\newcommand{\bfn}{{\boldsymbol n}}
\newcommand{\bfA}{{\boldsymbol A}}
\newcommand{\bfB}{{\boldsymbol B}}
\newcommand{\bfJ}{{\boldsymbol J}}
\newcommand{\bfS}{{\boldsymbol S}}
\newcommand{\saddle}{\mathrm{saddle}}
\newcommand{\can}{\mathrm{can}}
\newcommand{\const}{\mathrm{const.}}
\newcommand{\eq}{\,\mathrm{eq}}
\newcommand{\ext}{\,\mathrm{ext}}
\newcommand{\emem}{\,\mathrm{em}}
\newcommand{\inhom}{\,\mathrm{inhom}}
\newcommand{\Non}{\,\mathrm{Non}}
\newcommand{\wt}[1]{\widetilde{#1}}
\newcommand{\wh}[1]{\widehat{#1}}
\newcommand{\ch}[1]{\check{#1}}
\newcommand{\ol}[1]{\overline{#1}}
\newcommand{\ora}[1]{\overrightarrow{#1}}
\newcommand{\ii}{\imath}
\newcommand{\ic}{\iota}
\newcommand{\mbbP}{\mathbb{P}}
\newcommand{\mbbR}{\mathbb{R}}
\newcommand{\mbbN}{\mathbb{N}}
\newcommand{\mbbZ}{\mathbb{Z}}
\newcommand{\Leftrightup}[1]{\overset{\mathrm{#1}}{\Longleftrightarrow}}
\newcommand{\avg}[1]{\left\langle\,{#1}\, \right\rangle}
\newcommand{\step}{\lrcorner\hspace*{-0.55mm}\ulcorner}
\newcommand{\equp}[1]{\overset{\mathrm{#1}}{=}}
\newcommand{\nequp}[1]{\overset{\mathrm{#1}}{\tiny\neq}}
\newcommand{\equpg}[2]{\overset{\mathrm{#2}}{#1}}
\newcommand{\nin}{\in\hspace{-.78em}\setminus}
\newcommand{\fraX}{\mathfrak{X}}
\newcommand{\Gam}[1]{\Gamma{#1}}
\newcommand{\GamLamM}[1]{{\Gamma\Lambda^{{#1}}M}}
\newcommand{\GamLam}[2]{{\Gamma\Lambda^{{#2}}{#1}}}
\newcommand{\GTM}{\Gamma TM}
\newcommand{\GT}[1]{{\Gamma T{#1}}}
\newcommand{\inp}[2]{\left\langle\,  #1\, , \, #2\, \right\rangle}
\newcommand{\inpr}[2]{\left(\,  #1\, , \, #2\, \right)}
\newcommand{\rmt}[1]{{\mathrm{t}}({#1})}
\newcommand{\rmo}[1]{{\mathrm{o}}({#1})}
\newcommand{\paral}{/\hspace*{-2pt}/}

\newcommand{\Prooff}[1]{\noindent{\bf {#1}\,}}

\title{Liouville integrability of Beltrami-Maxwell fields with sources:\\ A geometric theory of magnetic surfaces  
}
\author{\large Shin-itiro GOTO 
\\
Center for  Mathematical Science and Artificial Intelligence,\\
Chubu University,\quad 
1200 Matsumoto-cho, Kasugai, Aichi 487-8501, Japan
}
\date{\today}
\maketitle
\begin{abstract}%
 We study classical Maxwell fields with non-trivial sources 
 in four-dimensional Minkowski spacetime.
 In particular, contact and symplectic geometric aspects are discussed. 
 As a point of departure,
 it is shown that a few classes of solutions with sources
 are constructed from the so-called Beltrami fields, where these
 fields are contact forms on three-dimensional Riemannian manifolds.
 In addition, several applications in these classes are provided.
  One is to give a solution to the London equations for superconductors.
 Another one is that, 
 if there is a non-trivial 
 conserved quantity along the magnetic vector field, 
 then  the Maxwell system can be viewed as a 
 Liouville integrable Hamiltonian system on a four-dimensional
 symplectic manifold, where this manifold  
 is obtained by a symplectization of the contact manifold. 
 As examples, Maxwell fields with 
 the Lundquist model, a restricted case of the ABC flow model, and so on,
 are shown to be integrable in the above sense.  
 
\end{abstract}%


\section{Introduction}
\label{section-introduction}
Maxwell's equations are a system of partial differential equations, 
and solutions are called Maxwell fields.
They describe electromagnetic fields under given conditions, and hence
they play a pivotal role in electromagnetism\,\cite{Landau1971,Jackson1998}.    
Despite its long history, 
further developments of the theory are
expected to clarify properties Maxwell fields
and are to facilitate their engineering applications\,\cite{Nakata2019}.   
One approach to develop the theory is based on 
formulations that enable to write these equations
in differential geometric languages\,\cite{Abraham1998,Heal2003,Kitano2012}. 
These formulations enable to apply various well-established techniques
in differential geometry for
analyzing electromagnetic fields.   
A well-known one is based on Riemannian geometry,
and it has contributed to 
gauge theory in high energy physics\,\cite{Nakahara,Frenkel,Benn1987}.  
There are other geometric approaches,
including contact geometry, and they are expected to
shed light on hidden aspects of Maxwell's
equations\,\cite{Dahl2004}. Here contact geometry is roughly speaking an odd
dimensional counterpart of
symplectic geometry\,\cite{McDuff,Hofer1994,Silva2008}. 
Its known applications in physics are found in 
the studies of thermodynamics\,\cite{Mrugala1990, Grmela2012, Goto2015,Bravetti2017,Schaft2018,Entov2026},
three-body problems\,\cite{Moreno2022,Frauenfelder2018}, 
general relativity\,\cite{Kozaki2022,Kozaki2024}, 
and in the studies of fluid mechanics\,\cite{Etnyre2000,Salas2024}. 

Maxwell's equations can be classified into two cases.
One is without any source term, and the other is with source terms, where 
source terms represent electric charge and current
for systems under consideration. 
Among a number of solutions to systems without source,
a particular class of Maxwell fields has been considered
due to its profound geometric nature.  
This class is such that electric fields are parallel to magnetic fields, and
is constructed from the eigenvalue problem for the curl operator, where
the eigen field of this eigenvalue problem is called the Beltrami field.
See \cite{Lakhtakia1994,Mochizuki2021} and
references therein for its brief history of the applications of these 
electromagnetic fields.
The Beltrami field can be written in the differential geometric language 
and then helps clarify properties of Maxwell fields.  
For instance, the existence of closed field lines at a fixed time
has been shown by employing this class of Maxwell fields\,\cite{Goto2024}.
Since this class of systems has no sources, it is natural to ask whether  
there is a counterpart Maxwell field for the class of systems that have
sources.

In this paper a class of Maxwell's equations with sources is considered, and 
the solutions are constructed with the Beltrami field.
Its basic properties are also shown.  
To be specific, let $(M,g)$ be a four-dimensional
pseudo Riemannian manifold, $\star$ a Hodge map associated with $g$, 
$F$ a two-form, $j_{4}$ a three-form called a source or a four-current, 
and $\varepsilon_{0}$ a positive constant expressing permittivity in vacuum.  
Maxwell's equations can be written in a Riemannian geometric 
description, and they form a system for $F$,
$\dr F=0$ and $\varepsilon_{0}\dr\star F=j_{4}$. 
Given $F$, one has the electromagnetic fields $\bm{B},\bm{e},\bm{D},\bm{h}$
that are compatible with Gibbs' notation.   
When $j_{4}$ is not identically zero, 
the system is called Maxwell's equations with a source or a Maxwell system
that has a source in this paper. To have such a solution with the Beltrami
field, one needs to consider a three-dimensional Riemannian manifold
$(N,\bm{g})$, so that $M=\mbbR\times N$ and
$g=-\dr x^{0}\otimes \dr x^{0}+\bm{g}$ with $x^{0}$ being
the coordinate of $\mbbR$. This coordinate $x^{0}$ is written as
$x^{0}=c_{0}t$ with $c_{0}$ being the speed of light in vacuum,
and $t\in\mbbR$ expresses time physically.
This $\bm{g}$ with a volume element $\sstar 1$ induces
the Hodge map $\sstar$. 
A rotational Beltrami field is a non-trivial 
one-form $\bm{\lambda}_{\kappa}$ 
satisfying $\sstar\bdr\bm{\lambda}_{\kappa}=\kappa\bm{\lambda}_{\kappa}$
and $\ii_{\partial/\partial x^{0}}\bm{\lambda}_{\kappa}=0$,  
where $\kappa\neq0$ is constant and $\bdr$ is the exterior derivative
that generates one-forms on $N$.
When a rotational Beltrami field is non-singular, this is called
a non-singular rotational Beltrami field. 
In this paper every object is assumed to be smooth, and $\bm{\lambda}_{\kappa}$
is assumed to be non-singular throughout. Hence,  
a non-singular rotational Beltrami field is abbreviated as
a rotational Beltrami field. Furthermore, a rotational Beltrami field
is abbreviated as a Beltrami field. 
The following is one of the main theorems in this paper.
\begin{Theorem}[Time dependent Beltrami-Maxwell fields with source 1]
\label{fact-introduction-1}
Let $\bm{\lambda}_{\kappa}$ be a  rotational Beltrami field on $N$, 
$f_{\ext}$ a function of $t=c_{0}^{-1}x^{0}\in\mbbR$, 
and
$$
j_{4}=\varepsilon_{0}(\sstar\bm{\lambda}_{\kappa})\wedge f_{\ext}\dr x^{0}.
$$ 
In addition, let $f$ be a solution to
the linear second order ordinary differential equation, 
\beq
\frac{\dr^{2}}{\dr t^{2}}f+c_{0}^{2}\kappa^{2}f
=-c_{0}^{2}f_{\ext}(t).
\label{ode1}
\eeq
Then, the two-form $F=\dr(f\bm{\lambda}_{\kappa})$ on $M$ is a solution to the
Maxwell system, 
$\dr F=0$ and $\varepsilon_{0}\dr \star F=j_{4}$.
\end{Theorem}
See Section\,\ref{section-fact-introduction-1-several} 
 for several solutions from
this solution. 
Similar to Theorem\,\ref{fact-introduction-1}, one has the following.
\begin{Theorem}
[Time dependent Beltrami-Maxwell fields with source 2]  
\label{fact-introduction-2}
Let $\bm{\lambda}_{\kappa}$ be a rotational Beltrami field on $N$, and  
$$
j_{4}=\varepsilon_{0}(\sstar\bm{\lambda}_{\kappa})\wedge f\dr x^{0}. 
$$
In addition, let $f$ be a solution to 
the linear second order ordinary differential equation,
\beq
\frac{\dr^{2}}{\dr t^{2}}f+c_{0}^{2}(\kappa^{2}+1)f
=0.
\label{ode2}
\eeq
Then, the two-form $F=\dr(f\bm{\lambda}_{\kappa})$ on $M$ 
is a solution to the Maxwell system,
$\dr F=0$ and $\varepsilon_{0}\dr \star F=j_{4}$.
\end{Theorem}
In this paper the Maxwell fields in Theorems\,\ref{fact-introduction-1}
and \ref{fact-introduction-2} are
called {\it Beltrami-Maxwell fields with sources}.  
As a corollary of Theorem\,\ref{fact-introduction-2},
the electromagnetic fields in Theorem \ref{fact-introduction-2}
are solutions to the London equations for superconductors.  
See Corollary\,\ref{fact-London-equations}.   
Although Maxwell fields without source has been described
in the language of contact geometry\,\cite{Goto2024,Dahl2008}, 
it was unclear to what extent this fact
could be generalized to cases with external sources. 
Theorems\,\,\ref{fact-introduction-1} and \,\ref{fact-introduction-2}
make this point clear.  

The solutions presented above are not only of interest
from a purely mathematical standpoint but also find 
applications in nature and engineering. For example,
time-independent Beltami-fields have been discussed in plasma physics.
This is because magnetic field configuration in interplanetary flux ropes
can be expressed as Beltrami fields\,\cite{Vandas2003}, and  
the Beltrami fields describe ordered structures of electromagnetic fields
in plasmas\,\cite{YoshidaZ1991Butsuri,YoshidaZ1991PTP}. 
One particular important object is the so-called magnetic surface\, 
\cite{YoshidaZ1991PTP,Vandas2017}.   
If there exists such a magnetic surface, then, roughly speaking,
the spatial structure of magnetic vector fields is ordered.
To speak of them properly, notions in symplectic and contact geometries 
are valuable. 
In this paper it is shown that this situation can be described
in the geometric language as follows.
\begin{Theorem}[Magnetic stream function and integrability]
 \label{fact-introduction-3}  
 If there is a magnetic stream function
 in the system of Theorem\,\ref{fact-introduction-1} and that of
 Theorem\,\ref{fact-introduction-2},  
 then it implies
  that Maxwell's systems are Liouville integrable on $M$.  
\end{Theorem}
See Theorem\,\ref{fact-magnetic-surface-integrability} for detail. 
Theorem\,\ref{fact-introduction-3} implies that the properties of 
magnetic fields associated with Maxwell fields with sources 
can be described in the language of symplectic and contact geometries. 
This means that the groundwork is laid for introducing topological methods 
into electromagnetism, 
similar to what has already been done in fluid dynamics\,\cite{Ghrist2001}. 
In addition, the integrability implies various properties in phase space, 
and enables to build perturbation theories\,\cite{Kozlov1983}.  
It then holds the promise of establishing new methods for 
analyzing electromagnetic fields, including magnetic surfaces, in the future. 

There are known Maxwell fields that are related to the solutions of this paper. 
The paper \cite{Flores2021} studied time-independent fields, whereas 
the present paper studies time-dependent fields. Hence, how to excite 
magnetic surfaces can be discussed with the solution given in this paper. 
Let us broaden the perspective of this paper, 
and consider fluid mechanics. 
The so-called Euler equation is one of the most important 
equations in fluid mechanics,  
and its class of solutions is written with  
the Beltrami field\,\cite{Cardona2019}. 
This indicates that the Beltrami field is a mathematical concept situated at 
the intersection of electromagnetism and fluid mechanics. 
This indication enables the mutual use of mathematical methods.
  
This article is organized as follows.
In Section\,\ref{section-preliminaries}
necessary background is briefly summarized, and   
the notations used in this paper are fixed, where such background is 
Riemannian and Lorentzian manifolds, symplectic and contact manifolds, 
Maxwell's equations,  and geometry related to rotational 
Beltrami fields. 
In Section\,\ref{section-Maxwell-fields-with-source}
Theorems\,\ref{fact-introduction-1} and \ref{fact-introduction-2} are proved,
and their consequences are provided.  
Such consequences include an application of 
the Weinstein conjecture and relations with 
the London equations for superconductors. 
In Section\,\ref{section-magnetic-surfaces} 
a theory of magnetic surfaces is proposed. 
Namely, a formal version of Theorem\,\ref{fact-introduction-3} is stated and 
proved. Explicit examples associated with this theorem are shown. 
In Section\,\ref{section-dicussion-conclusions}  
the present study is summarized, and  some potential future works are 
proposed.  
Section\,\ref{section-appendix} in appendix shows 
derivations of equations, explicit calculations of electromagnetic fields, 
and relevant examples discussed in the main text. 

\section{Preliminaries} 
\label{section-preliminaries}

In what follows underlining 
geometric objects are introduced, and notations
are fixed. Let $M$ be a four-dimensional manifold.
The space of vector fields on $M$ is denoted by $\GTM$, and  
the space of $k$-forms on $M$ is denoted by $\GamLamM{k}$, $k=0,\ldots,4$. 
The Lie derivative of a tensor field $\mathscr{T}$ along a vector field $X$
is denoted by $\cL_{X}\mathscr{T}$,
and the Cartan formula for a $k$-form $\alpha$ holds, that is, 
$\cL_{X}\alpha=(\dr \ii_{X}+\ii_{X}\dr)\alpha$,
where $\ii_{X}:\GamLamM{k}\to\GamLamM{k-1}$ denotes the
interior product with respect to $X\in\GTM$ 
and $\dr:\GamLamM{k}\to\GamLamM{k+1}$ denotes 
the exterior derivative. The Lie derivative of  $Y\in\GTM$ along $X$
is $\cL_{X}Y=[X,Y]$, where $[X,Y]=XY-YX$ is 
called the Lie bracket or the
commutator. It holds that
$\cL_{X}\ii_{Y}\alpha-\ii_{Y}\cL_{X}\alpha=\ii_{[X,Y]}\alpha, \forall \alpha\in\GamLamM{k},X,Y\in\GTM$. 
Let $t$ be the coordinate for $I\subset\mbbR$ called time, and
$c:I\to M$, $(t\mapsto c(t))$ be a curve. Then a differential of a tensor
field $\mathscr{T}$ with respect to $t$ is denoted by
$\cL_{\partial/\partial t}\mathscr{T}$. 
To discuss lower dimensional manifolds, 
let $N$ be a three-dimensional manifold. Then
$\GamLam{N}{k}, k=0,\ldots,3$ denotes the space of $k$-forms on $N$, and so on. 
The exterior derivative acting on $\GamLam{N}{k}$  is denoted by $\bdr$,  
when emphasizing the action is on $\GamLam{N}{k}$, rather than $\GamLamM{k}$.  
If a vector field $X$ does not vanish anywhere on some domain
under consideration, then $X$ is called {\it non-singular}.
If a one-form $\alpha$ does not vanish anywhere on some domain
under consideration, then $\alpha$ is also called {\it non-singular}.  

\subsection{Riemannian and Lorentzian manifolds}

Riemannian and Lorentzian geometries provide 
a natural description of the covariant form of Maxwell's equations 
as written in various textbooks\,\cite{Nakahara,Frenkel,Burton2024}.  
This description is also employed throughout this paper. 
In this subsection, Riemannian and Lorentzian geometries are 
summarized and the notations are fixed. 

Let $(M,g)$ be a four-dimensional Lorentzian manifold with
$M= I\times N$ with $N$ being a three-dimensional manifold and
$I\subset\mbbR$, where $g$ denotes a Lorentzian manifold on $M$. 
In addition, let $x=(x^{0},x^{1},x^{2},x^{3})$ be coordinates 
for $M$, and $c_{0}\in\mbbR$ a positive constant
that physically expresses the speed of light in vacuum, and then 
$t=x^{0}/c_{0}\in\mbbR$ physically expresses time. Since $g$ is Lorentzian,
it is written as 
$$
g=-\dr x^{0}\otimes \dr x^{0}+\bm{g}.
$$
This $\bm{g}$ is assumed to satisfy 
$\cL_{\partial/\partial x^{0}}\bm{g}=0$ throughout this paper. At a fixed 
$x^{0}$, $N_{x^{0}}$ denotes the hypersurface  in $M$, and  
$(N_{x^{0}},\bm{g})$ becomes a Riemannian manifold. 
For brevity, $(N_{x^{0}},\bm{g})$ is also denoted by $(N,\bm{g})$
when there is no risk of confusion.  
In a physics language, $M$ and $N$ are called {\it spacetime} and {\it space},
respectively.  
The metric can be written as 
$g_{0}=\sum_{a,b=0}^{3}\eta_{\,ab}\theta^{a}\otimes \theta^{b}$, where
 $\{\theta^{0},\ldots,\theta^{3}\}$
is an orthonormal coframe and $\eta_{ab}$ are elements of the diagonal matrix 
$(\eta_{ab})=\text{diag}\{-1,1,1,1\}$. 
Associated with $g_{0}$,    
$g_{0}^{-1}=\sum_{a,b=0}^{3}\eta^{\,ab}X_{a}\otimes X_{b}$ is introduced. Here
$(\eta^{ab})$ is the inverse of $(\eta_{ab})$ and $\{X_{0},\ldots,X_{3}\}$
is the dual basis of $\{\theta^{0},\ldots,\theta^{3}\}$,
$\theta^{a}(X_{b})=\delta_{b}^{a}$, where $\delta_{b}^{a}$ is the Kronecker delta  giving unity for $a=b$ and zero otherwise.
Similarly, $\bm{g}$ and $\bm{g}^{-1}$ can be written as $\bm{g}=\sum_{a=1}^{3}\bm{\sigma}^{a}\otimes\bm{\sigma}^{a}$ and $\bm{g}^{-1}=\sum_{a=1}^{3}e_{a}\otimes e_{a}$, respectively, where $\{\bm{\sigma}^{1},\bm{\sigma}^{2},\bm{\sigma}^{3}\}$ and $\{e_{1},e_{2},e_{3}\}$ are an orthonormal coframe and its dual  satisfying 
  $\bm{\sigma}^{a}(e_{b})=\delta_{b}^{a}$. 
The metric dual $\wt{\ }:\GTM\to\GamLamM{1}$ is introduced, and 
its inverse is also denoted $\wt{\ }:\GamLamM{1}\to\GTM$ by an abuse of 
notation:
$$
\wt{X}
=g(X,-),
\qquad
\wt{\alpha}
=g^{-1}(\alpha,-),
\qquad
\forall X\in \GTM,\quad
\forall \alpha\in \GamLamM{1}.
$$
Similarly, on $N$ the following are introduced:   
$$
\wt{X}^{\bm{g}}
=\bm{g}(X,-),\qquad
{\wt{\alpha}}^{\bm{g}}
=\bm{g}^{-1}(\alpha,-),\qquad
\forall X\in \GT{N},\
\forall \alpha \in \GamLam{N}{1}.
$$
These metric duals are isomorphisms, since $g$ and $\bm{g}$ are
non-degenerate. 
In addition, the Hodge maps associated with $g$ and $\bm{g}$ are denoted by
\beqa
\star
&:&\GamLamM{k}\to \GamLamM{4-k},\qquad
k=0,1,\ldots,4,
\non\\
\sstar
&:&
\GamLam{N}{k}\to \GamLam{N}{3-k},\qquad
k=0,1,2,3.
\non
\eeqa
There are several ways to define the Hodge map, and one of them is as follows. 
The action of $\star$  and that of $\sstar$  are determined by
\beqa
\star(\alpha_{1}\wedge\cdots\wedge\alpha_{r})
&=&\ii_{\wt{\alpha_{r}}}\cdots\ii_{\wt{\alpha_{1}}}\star 1,
\qquad 
\forall\ \alpha_{1},\cdots,\alpha_{r}\in\GamLamM{1}, 
\qquad
1\leq r\leq 4,
\non\\
\sstar(\beta_{1}\wedge\cdots\wedge\beta_{r})
&=&\ii_{\wt{\beta_{r}}}\cdots\ii_{\wt{\beta_{1}}}\sstar 1,
\qquad 
\forall\ \beta_{1},\cdots,\beta_{r}\in\GamLam{N}{1}, 
\qquad
1\leq r\leq 3,
\non
\eeqa
with fixed volume elements $\star 1$ and $\sstar 1$. 
It then follows for  $\alpha\in\GamLamM{k}$, 
$k=0,\ldots,3$ and  $\beta\in\GamLam{N}{k}$, $k=0,1,2$,   
that 
\beqa
\ii_{X}\star\alpha
&=&\star\left(\alpha\wedge\wt{X}\right),\qquad \forall\ X\in\GTM,
\non\\
\ii_{X}\sstar \beta
&=&\sstar \left(\beta\wedge\wt{X}^{\bm{g}}\right),\qquad \forall\ X\in\GT{N}.
\non
\eeqa
The relation between the volume elements $\star1$ and $\sstar 1$
is fixed in this paper as 
$$
\star 1
=\dr x^{0}\wedge\sstar 1,
$$
and 
$\sstar 1$ is typically of the form
$\bm{\sigma}^{1}\wedge\bm{\sigma}^{2}\wedge\bm{\sigma}^{3}$.
When $\bm{\sigma}^{a}=\bdr x^{a}, (a=1,2,3)$, it reduces to 
$\sstar 1=\bdr x^{1}\wedge\bdr x^{2}\wedge\bdr x^{3}$,
where $(x^{1},x^{2},x^{3})$ denotes coordinates of $N$.   
For the case 
$\bm{g}=\sum_{a=1}^{3}\bm{\sigma}^{a}\otimes\bm{\sigma}^{a}$ and 
$\sstar1=\bm{\sigma}^{1}\wedge\bm{\sigma}^{2}\wedge\bm{\sigma}^{3}$ 
at any point of $N$, it can be shown from
$$
\wt{\bm{\sigma}^{a}}^{\bm{g}}
=\bm{g}^{-1}(\bm{\sigma}_{a},-)
=e_{a},\qquad a=1,2,3,
$$
that
\begin{align}
&\sstar\bm{\sigma}^{1}
=\ii_{e_{1}}\sstar 1
=\bm{\sigma}^{2}\wedge \bm{\sigma}^{3},
\qquad
\sstar (\bm{\sigma}^{2}\wedge \bm{\sigma}^{3})
=\ii_{e_{3}}\ii_{e_{2}}\sstar 1
=\bm{\sigma}^{1},
\non\\
&\sstar \bm{\sigma}^{2}
=\ii_{e_{2}}\sstar 1
=\bm{\sigma}^{3}\wedge \bm{\sigma}^{1},
\qquad
\sstar(\bm{\sigma}^{3}\wedge \bm{\sigma}^{1})
=\ii_{e_{1}}\ii_{e_{3}}\sstar 1
=\bm{\sigma}^{2},
\non\\
&\sstar\bm{\sigma}^{3}
=\ii_{e_{3}}\sstar 1
=\bm{\sigma}^{1}\wedge \bm{\sigma}^{2},
\qquad
\sstar (\bm{\sigma}^{1}\wedge \bm{\sigma}^{2})
=\ii_{e_{2}}\ii_{e_{1}}\sstar 1
=\bm{\sigma}^{3},
\non
\end{align}
and
$$
\sstar(\bm{\sigma}^{1}\wedge\bm{\sigma}^{2}\wedge\bm{\sigma}^{3})
=1,
$$
from which
\beq
\sstar\sstar\beta
=\beta,\qquad
\forall \beta\in\GamLam{N}{k},\quad
k=0,\ldots,3.
\label{star3-star3}
\eeq
Similarly, it follows that 
\beq
\star\star\alpha
=(-1)^{k+1}\alpha,\qquad\forall \alpha\in\GamLamM{k},\quad
k=0,\ldots,4.
\label{star-star}
\eeq

The following lemma will be used in the next sections 
without mentioning explicitly.
\begin{Lemma}
\label{fact-star-decompositions-1}
\begin{enumerate}
\item
  For any $\alpha \in\GamLamM{k}$ and any $X\in\GTM$, $k=0,\ldots,4$,
  it follows that
  $$
  \star\ii_{X}\alpha
  =-(\star\alpha)\wedge\wt{X}.
  $$
\item
  It follows that 
  $$
  \star\sstar1
  =-\dr x^{0},\qquad\text{and}\qquad
  \star\dr x^{0}
  =-\sstar1.
  $$
\item
  For a one-form $\bm{\alpha}$
  that does not contain $\dr x^{0}$,
  $\ii_{\partial/\partial x^{0}}\bm{\alpha}=0$, it follows
  that 
  $$
  \star\sstar\bm{\alpha}
  =-\bm{\alpha}\wedge \dr x^{0}.
  $$
\end{enumerate}
\end{Lemma}
\begin{Proof}
\begin{enumerate}
\item
Write $\ii_{X}(\star\star\alpha)$ in the two different ways as
\begin{align}
  \ii_{X}(\star\star\alpha)
&=(-1)^{k+1}\ii_{X}\alpha
  \non\\
  &=\star\left[(\star\alpha)\wedge\wt{X}\right].
  \non
\end{align}
Since they are equal, one has that 
$$
(-1)^{k+1}\ii_{X}\alpha
=\star\left[(\star\alpha)\wedge\wt{X}\right].
$$
Applying $\star$ to the both sides above, one has 
from \fr{star-star} with $(\star\alpha)\wedge\wt{X}\in\GamLamM{4-k+1}$ that
$$
(-1)^{k+1}\star\ii_{X}\alpha
=\star\star\left[(\star\alpha)\wedge\wt{X}\right]
=(-1)^{4-k+1+1}(\star\alpha)\wedge\wt{X}.
$$
The obtained equation is equivalent to the desired equation. 

\item
Since
$\ii_{\wt{\dr x^{0}}}\dr x^{0}=g^{-1}(\dr x^{0},\dr x^{0})=-1$ and
$\star 1=\dr x^{0}\wedge\sstar1$, one has
$$
-\ii_{\wt{\dr x^{0}}}\star 1
=    -\ii_{\wt{\dr x^{0}}}(\dr x^{0}\wedge\sstar1)
=\sstar1.
$$
By applying  $\star$ to the both sides of the above equation
with \fr{star-star}, one obtains
$$
\star\sstar1
=-\star\ii_{\wt{\dr x^{0}}}\star 1
=-\star\star\dr x^{0}    
=-\dr x^{0}.
$$
This also yields
$$
\star\dr x^{0}
=-\star\star(\sstar1)
=-\sstar1.
$$
\item
Applying Items 1 and 2 with $\wt{\bm{\alpha}}=\wt{\bm{\alpha}}^{\bm{g}}$,
one has that
$$
\star\sstar\bm{\alpha}
=\star\ii_{\wt{\bm{\alpha}}}\sstar1
=-(\star\sstar 1)\wedge\bm{\alpha}
=\dr x^{0}\wedge\bm{\alpha}
=-\bm{\alpha}\wedge\dr x^{0}.
$$
\end{enumerate}
\qed
\end{Proof}

\subsection{Symplectic and contact manifolds}
\label{section-review-symplectic-contact}

Symplectic and contact geometries provide a natural description of
mechanical systems. In addition, 
these geometries have been employed to
clarify properties of some classes of solutions to Maxwell's equations
\,\cite{Goto2024,Dahl2008}. 
In this subsection, necessary background of 
symplectic and contact geometries are summarized, and notations are fixed.

One natural question about a given two-form $\omega$
on an even dimensional manifold
is whether $\omega$ is symplectic or not. 
Here $\omega$ on $M$ of dimension four is {\it symplectic} if
\beq
\dr\omega=0,\qquad\text{and}\qquad\omega\wedge\omega\neq 0
\label{conditions-to-be-symplectic}
\eeq
at any point of $M$.
If a symplectic form $\omega$ is provided on $M$, then $(M,\omega)$
is called a {\it symplectic manifold}. 
On symplectic manifolds, Hamiltonian vector fields  play
one of the central roles, where 
the Hamiltonian vector field $X_{\cal H}$ associated with
a given function ${\cal H}$ on $M$ 
is the uniquely determined vector field satisfying 
$$
\ii_{X_{\cal H}}\omega
=-\dr{\cal H}. 
$$
Define the {\it Poisson bracket} for two functions on $M$ as
$\GamLamM{0}\times\GamLamM{0}\to\GamLamM{0}$ with 
$$
\{{\cal H},{\cal H}^{\prime}\}_{\omega}
:=\omega(X_{{\cal H}},X_{{\cal H}^{\prime}}),\quad
\forall {\cal H},{\cal H}^{\prime}\in\GamLamM{0}.
$$
This can be written with the Hamiltonian vector field $X_{\cal H}$ as 
$$
\{{\cal H},{\cal H}^{\prime}\}_{\omega}
=-\dr {\cal H}(X_{{\cal H}^{\prime}})
=-X_{{\cal H}^{\prime}}{\cal H}
=X_{{\cal H}}{\cal H}^{\prime}.
$$
On a symplectic manifold of dimension four $(M,\omega)$, 
the Hamiltonian system with the pair of functions 
$({\cal H},{\cal H}^{\prime})$ on $M$ is called {\it Liouville integrable} 
if (i) $\dr {\cal H}\wedge\dr {\cal H}^{\prime}\neq 0$ almost everywhere, and
(ii) they commute with respect to the Poisson bracket:
$$
\{{\cal H},{\cal H}^{\prime}\}_{\omega}
=0.
$$
The functions ${\cal H}$ and ${\cal H}^{\prime}$ that satisfy the above equation 
are called {\it Poisson commuting functions}, and they are also said to be
in {\it involution}. 

One natural question about a given one-form $\bm{\lambda}$
on an odd dimensional manifold
is whether $\bm{\lambda}$ is contact or not. 
Here $\bm{\lambda}$ on $N$ of dimension three is  {\it contact} if
\beq
\bm{\lambda}\wedge\bdr\bm{\lambda}\neq 0, 
\label{conditions-to-be-contact}
\eeq
at any point of $N$.
If a contact form is provided on $N$, then $(N,\ker\bm{\lambda})$ 
is called a {\it contact manifold}, where
$\ker\bm{\lambda}:=\{X\in\GT{N} | \bm{\lambda}(X)=0 \}$. 
Associated with a given $\bm{\lambda}$,
there is a unique vector field called the {\it Reeb vector field} $R$
such that
$$
\ii_{R}\bm{\lambda}
=1,\qquad \text{and}\qquad
\ii_{R}\bdr\bm{\lambda}
=0.
$$

Given a contact manifold, 
there is a way to introduce a symplectic manifold. 
The following way is called a {\it symplectization}. 
To introduce a four-dimensional symplectic manifold
with the contact manifold $(N,\ker\bm{\lambda})$, 
let $\vartheta$ be a function on $I\subset\mbbR$
such that $\vartheta$ and $\dr\vartheta$ do not vanish anywhere on $I$. Then 
$(I\times N,\dr(\vartheta \bm{\lambda}))$ is a symplectic manifold.
Indeed, the two-form $\dr(\vartheta \bm{\lambda})$ is a symplectic form due to
$$
\dr(\vartheta\bm{\lambda})
\wedge\dr(\vartheta\bm{\lambda})
=2\vartheta\dr\vartheta\wedge\bm{\lambda}\wedge\bdr\bm{\lambda}
\neq 0,
$$
at any point of $I\times N$. In the literature $\vartheta(t)=\exp(t)$ with $I=\mbbR$ is often used. 

\subsection{Maxwell's equations}
\label{section-preliminaries-Maxwell-equation}
Maxwell's equations together with given various data describe 
electromagnetic fields\,\cite{Jackson1998}.
In the covariant form of electromagnetism on spacetime $(M,g)$, 
electromagnetic fields are written in terms of a pair of two-forms
in general. In simple cases, however, a one two-form is enough, and it is
denoted by $F$. 
To express a source $j_{4}\in \GamLamM{3}$, called a {\it four-current},
is introduced. 
In this case, Maxwell's equations are written as   
\beq
\dr F
=0,\qquad\text{and}\qquad
\varepsilon_{0}\dr \star F
=j_{4},
\label{Maxwell-equations-vacuum}
\eeq
where $\varepsilon_{0}>0$ is the constant called the
{\it electric permittivity} in vacuum.
For later use let $\mu_{0}>0$ be the constant called the 
{\it permeability} in vacuum. 
There is a relation among $c_{0},\varepsilon_{0}$ and $\mu_{0}$, and it is
$$
c_{0}
=\frac{1}{\sqrt{\varepsilon_{0}\mu_{0}}}.
$$
The two-form $F$ is called the {\it Maxwell two-form}. 
This $F$ is also called the {\it (electromagnetic) field strength}. 

Suppose that $F\in\GamLamM{2}$ is decomposed with respect to $\dr x^{0}$ as
\beq
F=-c_{0}\bm{B}-\bm{e}\wedge\dr x^{0},
\label{F0}
\eeq
and introduce
\beq
\bm{D}=\varepsilon_{0}\sstar\bm{e},\qquad
\bm{h}=\mu_{0}^{-1}\sstar\bm{B},
\label{Maxwell-constitutive-relations-vacuum-decomposed}
\eeq
where  
$\bm{B},\bm{D}\in\GamLamM{2}$ and $\bm{e},\bm{h}\in\GamLamM{1}$  
satisfy
$$
\ii_{\partial/\partial x^{0}}\bm{B}
=\ii_{\partial/\partial x^{0}}\bm{D}
=\ii_{\partial/\partial x^{0}}\bm{e}
=\ii_{\partial/\partial x^{0}}\bm{h}
=0.
$$
It then follows that one can extract $\bm{e}$ and $\bm{h}$ from $F$ as 
\beq
\bm{e}
=\ii_{\partial/\partial x^{0}}F,\qquad\text{and}\qquad
\bm{h}
=-\varepsilon_{0}c_{0}\ \ii_{\partial/\partial x^{0}}\star F,
\label{e-h-from-F}
\eeq
(see Section \ref{section-derivation-e-h-from-F} for the derivations). 

Suppose that $j_{4}\in\GamLamM{3}$ is decomposed with respect to $\dr x^{0}$ as 
\beq
j_{4}=-\rho\sstar1+c_{0}^{-1}\bm{J}\wedge\dr x^{0},
\label{j0}
\eeq
where $\rho$ is a nought-form, and $\bm{J}$  is a two-form satisfying
$$
\ii_{\partial/\partial x^{0}}\bm{J}
=0.
$$
For later use, the one-form $\bm{j}$ is introduced with $\bm{J}$ as
$$
\bm{j}=\sstar\bm{J}, \qquad\text{from which}\quad
\ii_{\partial/\partial x^{0}}\bm{j}
=0.
$$
From \fr{j0}, it follows that  
$$
\dr j_{4}
=c_{0}^{-1}\left(\dot{\rho}\sstar1
+\bdr\bm{J}\right)\wedge \dr x^{0},
$$
where $\dot{\rho}:=\partial\rho/\partial t$. Note that
$$
\dot{\rho}\dr t
=c_{0}^{-1}\dot{\rho}\dr x^{0},
$$
see also \fr{dot-alpha-is-Lie} below.  
Hence the equation $\dr j_{4}=0$ leads to the 
{\it continuity equation}:
\beq
\bm{\dr J}+\dot{\rho}\sstar1
=0.
\label{continuity-equation-dj=0}
\eeq

In physics, these one-form and two-form fields are termed 
as shown in Table\,\ref{table-forms-names}.
\begin{table}[htb]
\caption{Tensors for Maxwell's equations}
\label{table-forms-names}
\begin{center}
  \begin{tabular}{|c|c|}
    \hline
    $F\in\GamLamM{2}$& Maxwell two-form\\
    \hline
    $\bm{j}\in\GamLamM{3}$& source three-form\\
    \hline
    $\bm{B}\in\GamLamM{2}$&magnetic flux field\\
    \hline
    $\bm{D}\in\GamLamM{2}$&electric displacement field\\
    \hline
    $\bm{e}\in\GamLamM{1}$&electric field\\
    \hline
    $\bm{h}\in\GamLamM{1}$&magnetic field\\
    \hline
    $\rho\sstar1\in\GamLamM{0}$& charge three-form\\
    \hline
    $\bm{J}\in\GamLamM{2}$& current two-form\\
    \hline
  \end{tabular}
\end{center}    
\end{table}  

Throughout this paper we focus on the following:
\begin{Definition}
\label{definition-Maxwell's-equations}
{\it Maxwell's equations
  in vacuum with a source $j_{4}$ on a Riemannian manifold $(M,g)$} are 
\fr{Maxwell-equations-vacuum}:
$$
\dr F
=0,\qquad\text{and}\qquad
\varepsilon_{0}\dr\star F
=j_{4}.
$$
A solution $F$, or a set of fields
$\bm{e},\bm{h},\bm{D},\bm{B}$, are called {\it Maxwell fields}.
In particular, $\bm{e}$ and $\bm{h}$
are called {\it electromagnetic one-forms},  
$\wt{\bm{e}}^{\bm{g}}$ and $\wt{\bm{h}}^{\bm{g}}$ are called
{\it electromagnetic vector fields}.
If fields depend on $x^{0}$, then they are called {\it time-dependent}.
Otherwise they are called {\it time-independent}. 
\end{Definition} 
Applying the exterior derivative of the both sides of
\fr{Maxwell-equations-vacuum}, one has that $\dr j_{4}=0$.
This leads to the continuity equation \fr{continuity-equation-dj=0}.  

To rewrite Maxwell's equations in a decomposed form, extend $\bdr$ such that 
$$
\bdr:\GamLam{M}{k}\to
\GamLam{M}{k+1},\qquad k=0,1,\ldots,4,
$$
such that\,\cite{Frenkel} 
$$
\dr=
\bdr+\dr x^{0}\wedge \cL_{\partial/\partial x^{0}}.
$$
For example, if $\Upsilon$ is a function of $x=(x^{0},\ldots,x^{3})$, then
$$
\dr\Upsilon
=\frac{\partial\Upsilon}{\partial x^{0}}\dr x^{0}+\bdr \Upsilon,\qquad
\bdr \Upsilon
=\sum_{a=1}^{3}\frac{\partial \Upsilon}{\partial x^{a}}\bdr x^{a}.
$$
In addition, the abbreviation for the Lie derivative of any $k$-form
$\alpha$ on $M$, $k=0,\ldots,4$, with respect to $\partial/\partial t$   
is introduced as 
\begin{align}
\dot{\alpha}
=\cL_{\partial/\partial t}\alpha.
\label{dot-alpha-is-Lie}
\end{align}

Then Maxwell's equations \fr{Maxwell-equations-vacuum} for $F$  
with \fr{Maxwell-constitutive-relations-vacuum-decomposed}
are decomposed into the equations for $\bm{B},\bm{D},\bm{e},\bm{h}$ as 
\beq
\dot{\bm{B}}
=-\bdr\bm{e},\qquad
\bdr\bm{B}
=0,\qquad
\bdr\bm{D}
=\rho\sstar1,\qquad
\dot{\bm{D}}
=\bdr\bm{h}-\bm{J},
\label{Maxwell-decomposed-vacuum}
\eeq
(see Section\,\ref{section-appendix-Maxwell-decomposed} 
for their derivations). This decomposition \fr{Maxwell-decomposed-vacuum} 
is compatible with
Gibb's vector notation.
In \fr{Maxwell-decomposed-vacuum}, $\rho$ and $\bm{J}$ should be chosen so that
the continuity equation \fr{continuity-equation-dj=0} holds.  
 
The concept of energy is important in physics. 
For Maxwell fields, energy density forms are defined.  
They are the three-forms
\beq
\bm{\cE}_{\e}
:=\frac{1}{2}\bm{e}\wedge\bm{D},\qquad
\bm{\cE}_{\m}
:=\frac{1}{2}\bm{h}\wedge\bm{B},\qquad\text{and}\qquad
\bm{\cE}_{\emem}
:=\bm{\cE}_{\e}+\bm{\cE}_{\m},
\label{energy-densities}
\eeq
where $\bm{\cE}_{\emem}$ is called the
{\it total electromagnetic density three-form} in this paper.
The integral of $\bm{\cE}_{\emem}$ over a non-empty
region $N^{\prime}\subset N$ is called the {\it total electromagnetic energy}.  
In addition, the two-form defined as 
\beq
\bm{\cI}
:=\bm{e}\wedge\bm{h},
\label{Poynting-2-form}
\eeq
is called the {\it Poynting two-form} in this paper, and this two-form 
corresponds to the Poynting vector 
in classical electromagnetism\,\cite{Jackson1998}.  

\subsection{Rotational Beltrami one-form}
Rotational Beltrami fields will be employed to construct solutions to
Maxwell's equations in Section \ref{section-Maxwell-fields-with-source}. 
There are several ways to introduce rotational Beltrami fields in general. 
In this paper rotational Beltrami fields
are one-forms on $(N,\bm{g})$, and these forms can be  
transformed into vectors uniquely 
with the use of the metric dual with respect to $\bm{g}$. 
The details of these are discussed below.   
\begin{Definition}
\label{definition-Beltrami-1-form}
Let $(N,\bm{g})$ be a three-dimensional Riemannian manifold.
A one-form $\bm{\lambda}_{\kappa}$ on $N$ labeled by a real constant $\kappa$  
is a {\it rotational Beltrami one-form} if $\bm{\lambda}_{\kappa}$ does not
identically vanish and it satisfies 
\beq
\sstar\bdr\bm{\lambda}_{\kappa}
=\kappa \bm{\lambda}_{\kappa},\qquad \kappa\neq 0. 
\label{Beltrami-1-form-0}
\eeq
Given such a $\bm{\lambda}_{\kappa}$,
\beq
Z_{\kappa}
:=\bm{g}^{-1}(\bm{\lambda}_{\kappa},-)\quad \in\GT{N},
\label{Z-kappa}
\eeq
is called a {\it rotational Beltrami vector field}. 
\end{Definition}
The $\bm{\lambda}_{\kappa}$ in \fr{Beltrami-1-form-0} can be viewed as 
a non-trivial solution to the eigenvalue problem 
associated with the operator $\sstar\bdr$ acting on a one-form.
Throughout this paper, rotational Beltrami one-forms and
vector fields are assumed to be non-singular,
meaning that these never vanish on $N$.    
See Section \ref{section-example-Beltrami-form} for several examples of
rotational Beltrami one-form. 

There are several properties of \fr{Beltrami-1-form-0},
including the following lemma.
This Lemma will be used without mentioning explicitly.
\begin{Lemma}
\label{fact-Beltrami-field-general-1}
Let $\bm{\lambda}_{\kappa}$ be a one-form that satisfies
\fr{Beltrami-1-form-0} on $N$. Then the following hold. 
\begin{enumerate}
\item
$$
\bdr\sstar\bm{\lambda}_{\kappa}
=0.
$$
\item
$$
\bm{\lambda}_{\kappa}\wedge\sstar\bm{\lambda}_{\kappa}
=\bm{g}^{-1}(\bm{\lambda}_{\kappa},\bm{\lambda}_{\kappa})\sstar1.
$$
\item
(\cite{Etnyre2000}). The one-form $\bm{\lambda}_{\kappa}$ is a contact form.   
\end{enumerate}
\end{Lemma}
\begin{Proof}
\begin{enumerate}
\item
Straightforward calculations with \fr{star3-star3} yield 
$$
\bdr\sstar\bm{\lambda}_{\kappa}
=    \bdr(\kappa^{-1}\bdr\bm{\lambda}_{\kappa})
=0.
$$
\item
Applying $\ii_{Z_{\kappa}}$ to the both 
sides of $0=\bm{\lambda}_{\kappa}\wedge\sstar1$, one has that 
$$
0=\ii_{Z_{\kappa}}(\bm{\lambda}_{\kappa}\wedge\sstar1)
=\bm{g}^{-1}(\bm{\lambda}_{\kappa},\bm{\lambda}_{\kappa})\sstar1
-\bm{\lambda}_{\kappa}\wedge\sstar\bm{\lambda}_{\kappa},
$$
from which the required equation is obtained.
\item
Since $\bm{\lambda}_{\kappa}$ is assumed to be non-singular
throughout this paper and Item 2 holds,  
$\bm{\lambda}_{\kappa}$ is a contact form on $N$. 
\end{enumerate}
\qed
\end{Proof}

One can consider a weak version of Beltrami one-form such that
$$
\sstar\bdr\bm{\lambda}
=f \bm{\lambda},
$$
where $f$ is a function on $N$.  
Although sometimes this weak version is discussed
in the literature\,\cite{Benn1996,Enciso2016},  
this weak version will not intensively be considered in this paper. 

\section{Solutions to Maxwell fields with sources }
\label{section-Maxwell-fields-with-source}

In this section Theorems\,\ref{fact-introduction-1} and
\ref{fact-introduction-2} are proved.
Then basic properties of the Maxwell fields are discussed. 

\subsection{Proof of Theorem\,\ref{fact-introduction-1}}

It is now a position to prove Theorem\,\ref{fact-introduction-1}, in which
the source $j_{4}$ is specified by \fr{j0} with 
$$
\bm{J}=\varepsilon_{0}c_{0}f_{\ext}\sstar\bm{\lambda}_{\kappa},\qquad
\text{and}\qquad
\rho=0.
$$
\Prooff{Proof of Theorem\,\ref{fact-introduction-1}}
The strategy for proving this theorem is to substitute the ansatz
into Maxwell's equations with the source.
Recall that $F=\dr(f\bm{\lambda}_{\kappa})$, where
$f$ only depends on $t=c_{0}^{-1}x^{0}$ and $\dot{f}=\dr f/\dr t$.   

1. Equation $\dr F=0$ is verified as
$$
\dr F
=\dr\dr(f\bm{\lambda}_{\kappa})
=0.
$$ 

2. Equation $\varepsilon_{0}\dr \star F=j_{4}$ is verified as follows.
Firstly, the left hand side is calculated.
It follows from \fr{Beltrami-1-form-0} that
\begin{align}
F&=\dr (f\bm{\lambda}_{\kappa})
\non\\
&=c_{0}^{-1}\dot{f}\dr x^{0}\wedge\bm{\lambda}_{\kappa}+f\bdr\bm{\lambda}_{\kappa},
\non\\
\star F
&=c_{0}^{-1}\dot{f}\star(\dr x^{0}\wedge\bm{\lambda}_{\kappa})
+f\kappa\star(\sstar\bm{\lambda}_{\kappa}).
\non
\end{align}
To reduce the above expression of $\star F$, using
\begin{align}
&\star(\dr x^{0}\wedge\bm{\lambda}_{\kappa})
=\ii_{Z_{\kappa}}\ii_{\wt{\dr x^{0}}}\star 1
=\ii_{Z_{\kappa}}(-\sstar1)
=-\sstar\bm{\lambda}_{\kappa},
\non\\
&\star\sstar\bm{\lambda}_{\kappa}
=\dr x^{0}\wedge\bm{\lambda}_{\kappa}, 
\non
\end{align}
one has 
\begin{align}
\star F
&=c_{0}^{-1}\dot{f}(-\sstar\bm{\lambda}_{\kappa})
+\kappa f\dr x^{0}\wedge \bm{\lambda}_{\kappa},
\non\\
\dr \star F
&=-c_{0}^{-2}\ddot{f}\dr x^{0}\wedge\sstar\bm{\lambda}_{\kappa}
-\kappa f\dr x^{0}\wedge\bdr\bm{\lambda}_{\kappa}
\non\\
&=-(c_{0}^{-2}\ddot{f}+\kappa^{2}f)(\sstar\bm{\lambda}_{\kappa})\wedge\dr x^{0}.
\non
\end{align}
Then, one substitutes $c_{0}^{-2}\ddot{f}+\kappa^{2}f=-f_{\ext}$ into
$\varepsilon_{0}\dr\star F$,  and recalls  
$j_{4}=\varepsilon_{0}f_{\ext}(\sstar\bm{\lambda}_{\kappa})\wedge\dr x^{0}$.
These yield 
$$
\varepsilon_{0}\dr \star F
=\varepsilon_{0}f_{\ext}(\sstar\bm{\lambda}_{\kappa})\wedge\dr x^{0}
=j_{4}.
$$
Thus the theorem is proved.
\qed

The fields $\bm{e}$ and $\bm{h}$ can be extracted from $F$ as \fr{e-h-from-F}.
In addition, $\bm{D}$ and $\bm{B}$ can be obtained from
\fr{Maxwell-constitutive-relations-vacuum-decomposed}. They are 
\begin{align}
&\bm{e}
=\ii_{\partial/\partial x^{0}}F
=c_{0}^{-1}\dot{f}\bm{\lambda}_{\kappa},
\label{e-from-f}\\
&\bm{h}
=-\varepsilon_{0}c_{0}\ \ii_{\partial/\partial x^{0}}\star F
=-\varepsilon_{0}c_{0}\ \kappa f\bm{\lambda}_{\kappa},  
\label{h-from-f}\\
&\bm{D}
=\varepsilon_{0}\sstar\bm{e}
=\varepsilon_{0}c_{0}^{-1}\dot{f}\sstar\bm{\lambda}_{\kappa},
\non\\
&\bm{B}
=\mu_{0}\sstar\bm{h}
=-c_{0}^{-1}\kappa f\sstar\bm{\lambda}_{\kappa}.
\non
\end{align}
Observe from these expressions,
the Poynting two-form \fr{Poynting-2-form} vanishes identically at any point
of $N$ for all $t$, 
$$
\bm{\cI}
=\bm{e}\wedge\bm{h}
=-\varepsilon_{0}\kappa f\dot{f}{\bm\lambda}_{\kappa}\wedge
{\bm\lambda}_{\kappa}
=0.
$$
The energy densities \fr{energy-densities} are then written,
with emphasis on the $f$ dependence, as 
\begin{align}
&\bm{\cE}_{\e}^{f}
=\frac{1}{2}\bm{e}\wedge\bm{D}
=\frac{\varepsilon_{0}}{2}(c_{0}^{-1}\dot{f})^{2}
\bm{\lambda}_{\kappa}\wedge\sstar\bm{\lambda}_{\kappa},
\non\\
&\bm{\cE}_{\m}^{f}
=\frac{1}{2}\bm{h}\wedge\bm{B}
=\frac{\varepsilon_{0}}{2}(\kappa f)^{2}
\bm{\lambda}_{\kappa}\wedge\sstar\bm{\lambda}_{\kappa},
\non\\
&\bm{\cE}_{\emem}^{f}
=\bm{\cE}_{\e}^{f}+\bm{\cE}_{\m}^{f}
=\frac{\varepsilon_{0}}{2}\left\{
(c_{0}^{-1}\dot{f})^{2}+(\kappa f)^{2}
\right\}
\bm{\lambda}_{\kappa}\wedge\sstar\bm{\lambda}_{\kappa}.
\label{total-energy-3-form-theorem1}
\end{align}
Note that since $\bm{\lambda}_{\kappa}$
is assumed to be non-singular throughout this paper,  
the three-form
$\bm{\lambda}_{\kappa}\wedge\sstar\bm{\lambda}_{\kappa}=\bm{g}^{-1}(\bm{\lambda}_{\kappa},\bm{\lambda}_{\kappa})\sstar1$ is a volume-form.  
Note also that
$\cE_{\emem}^{f_{\I}+f_{\II}}\neq \cE_{\emem}^{f_{\I}}+\cE_{\emem}^{f_{\II}}$, where
$f_{\I}$ and $f_{\II}$ are solutions to \fr{ode1}:
$$
\cE_{\emem}^{f_{\I}+f_{\II}}
-(\cE_{\emem}^{f_{\I}}+\cE_{\emem}^{f_{\II}})
=\varepsilon_{0}\left(
c_{0}^{-2}\dot{f}_{\I}\dot{f}_{\II}+\kappa^{2}f_{\I}f_{\II}
\right)
\bm{\lambda}_{\kappa}\wedge\sstar\bm{\lambda}_{\kappa}.
$$

\subsection{Geometrical methods related to Theorem\,\ref{fact-introduction-1}}

As discussed in the literature, the rotational Beltrami field leads to
a contact form on a three-dimensional manifold\,\cite{Etnyre2000}
(Item 3 of Lemma\,\ref{fact-Beltrami-field-general-1}). 
It is then expected that known theorems in contact geometry  
can be applied to electromagnetic fields generated by the rotational
Beltrami field\,\cite{Dahl2004,Goto2024}.  In this subsection 
relations and consequences concerned with
Theorem\,\ref{fact-introduction-1} are discussed from 
a viewpoint of symplectic and contact geometries.

One natural question from the viewpoint of symplectic geometry is whether $F$
in Theorem\,\ref{fact-introduction-1} is symplectic or not.
The following is an answer.
\begin{Proposition}[Maxwell form is symplectic 1] 
\label{fact-F-symplectic}
Let $F$ be the solution in Theorem\,\ref{fact-introduction-1}. 
 When $f\dot{f}\neq 0$, this $F$ is symplectic.
\end{Proposition}
\begin{Proof}
To prove this, one shows that $\dr F=0$ and $F\wedge F\neq 0$
whenever $f\dot{f}\neq 0$. 
Firstly, it follows that $\dr F=0$.
Secondly, straightforward calculations show that 
$$
F\wedge F
=2c_{0}^{-1}\kappa f\dot{f}
\dr x^{0}\wedge\bm{\lambda}_{\kappa}\wedge\sstar\bm{\lambda}_{\kappa},
$$
where $\bm{\lambda}_{\kappa}\wedge\sstar\bm{\lambda}_{\kappa}\neq 0$ holds 
at any point of $N$.
When $f\dot{f}\neq 0$, $F$ satisfies $F\wedge F\neq 0$.
Hence $F$ satisfies the two conditions in \fr{conditions-to-be-symplectic}.
\qed
\end{Proof}
One natural question from the viewpoint of contact geometry 
is whether $\bm{e}$ in \fr{e-from-f} and $\bm{h}$ in \fr{h-from-f} 
are contact or not. The following is an answer.
\begin{Proposition}[Electromagnetic one-forms are contact 1]
\label{fact-e-h-contact}
On the hypersurface $\{\{t\}\times N |\dot{f}(t)\neq0\}\subset M$,
the $\bm{e}$ in \fr{e-from-f} is contact. In addition, 
on the hypersurface $\{\{t\}\times N| f(t)\neq0\}\subset M$, 
the $\bm{h}$ in \fr{h-from-f} is contact.   
\end{Proposition}
\begin{Proof}
To prove the statement on $\bm{e}$,
one calculates $\bm{e}\wedge\bdr\bm{e}$. It follows from
\fr{e-from-f} and \fr{Beltrami-1-form-0} that  
\begin{align}
\bm{e}\wedge\bdr\bm{e}
&= (c_{0}^{-1}\dot{f}\bm{\lambda}_{\kappa})
\wedge\bdr( c_{0}^{-1}\dot{f}\bm{\lambda}_{\kappa})
\non\\
&=(c_{0}^{-1}\dot{f})^{2}\bm{\lambda}_{\kappa}\wedge\bdr\bm{\lambda}_{\kappa}.
\non
\end{align}
Recall from  Item 3 of Lemma\,\ref{fact-Beltrami-field-general-1} that
$\bm{\lambda}_{\kappa}$ is contact,
$\bm{\lambda}_{\kappa}\wedge\bdr\bm{\lambda}_{\kappa}\neq0$. 
This and $(c_{0}^{-1}\dot{f})^{2}\neq 0$ yield 
$\bm{e}\wedge\bdr\bm{e}\neq 0$ on $\{\{t\}\times N|\dot{f}(t)=0\}$.
Thus, $\bm{e}$ is contact.   
A similar argument holds for $\bm{h}$, which yields
$\bm{h}\wedge\bdr\bm{h}\neq 0$ on $\{\{t\}\times N|f(t)=0\}$.
\qed
\end{Proof}

Note that the symplectic form $F$ in Proposition\,\ref{fact-F-symplectic} is a 
symplectization of the contact form $\bm{\lambda}_{\kappa}$ on $N$.
Associated with the contact form $\bm{\lambda}_{\kappa}$,
there is the unique Reeb vector field $R_{\kappa}$ such that
$$
\ii_{R_{\kappa}}\bm{\lambda}_{\kappa}
=1,\qquad\text{and}\qquad
\ii_{R_{\kappa}}\bdr\bm{\lambda}_{\kappa}
=0.
$$
The vector field
$\varphi_{*}R_{\kappa}\in \GTM$ is introduced from $R_{\kappa}\in \GT{N}$,
where $\varphi:N\to M$ is the embedding and $\varphi_{*}:\GT{N}\to\GTM$ 
denotes its push-forward map. For brevity, $\varphi_{*}R_{\kappa}$ is
often written simply as $R_{\kappa}$ when there is no risk of confusion. 
The following is an explicit expression of the Reeb vector field
$R_{\kappa}$ associated with $\bm{\lambda}_{\kappa}$
satisfying \fr{Beltrami-1-form-0}.
\begin{Lemma}
\label{fact-Reeb-from-Beltrami-form}
  The Reeb vector field
$R_{\kappa}$ associated with the contact form $\bm{\lambda}_{\kappa}$ 
satisfying \fr{Beltrami-1-form-0} is
$$
R_{\kappa}
=\frac{1}{\bm{g}^{-1}(\bm{\lambda}_{\kappa},\bm{\lambda}_{\kappa})}Z_{\kappa},
$$
where $Z_{\kappa}$ has been defined in \fr{Z-kappa}.
\end{Lemma}
\begin{Proof}
Given any contact form, it is known that 
the Reeb vector field exists uniquely. This vector is written  as $R_{\kappa}$.
Firstly, applying $\ii_{R_{\kappa}}$ to the identity 
$$
\bm{\lambda}_{\kappa}\wedge\bdr\bm{\lambda}_{\kappa}
=\kappa\bm{\lambda}_{\kappa}\wedge\sstar\bm{\lambda}_{\kappa}
=\kappa\bm{g}^{-1}(\bm{\lambda}_{\kappa},\bm{\lambda}_{\kappa})\sstar1,
$$
with $\ii_{R_{\kappa}}\bm{\lambda}_{\kappa}=1$
and $\ii_{R_{\kappa}}\bdr\bm{\lambda}_{\kappa}=0$, one has that
$$
\bdr\bm{\lambda}_{\kappa}
=\kappa \bm{g}^{-1}(\bm{\lambda}_{\kappa},\bm{\lambda}_{\kappa})
\sstar\wt{R_{\kappa}}^{\bm{g}}.
$$
Secondly, applying $\sstar$ to the both sides of the above equation,
one has that 
$$
\wt{R_{\kappa}}^{\bm{g}}
=\frac{\kappa^{-1}}{\bm{g}^{-1}(\bm{\lambda}_{\kappa},\bm{\lambda}_{\kappa})}
\sstar\bdr\bm{\lambda}_{\kappa}
=\frac{1}{\bm{g}^{-1}(\bm{\lambda}_{\kappa},\bm{\lambda}_{\kappa})}
\bm{\lambda}_{\kappa}.
$$
Finally,
taking the metric dual with respect to $\bm{g}$ of the obtained equation, 
one arrives at the desired equation. 
\qed
\end{Proof}
Since the Reeb vector field plays various roles in contact and symplectic
geometries\,\cite{McDuff,Hofer1994,Silva2008}, 
it is worth discussing what role the Reeb vector field plays. 
The following shows that the electromagnetic fields can be written
in terms of $R_{\kappa}$. 
\begin{Proposition}[Relations between the Reeb vector and electromagnetic fields 1]
\label{fact-e-h-by-R}
The vector fields $\wt{\bm{e}}^{\bm{g}}$ and $\wt{\bm{h}}^{\bm{g}}$, given by
\fr{e-from-f} and \fr{h-from-f}, are written as
\begin{align}
\wt{\bm{e}}^{\bm{g}}
&=c_{0}^{-1}\dot{f}\bm{g}^{-1}(\bm{\lambda}_{\kappa},\bm{\lambda}_{\kappa})
R_{\kappa},
\non\\
\wt{\bm{h}}^{\bm{g}}
&=-\,\varepsilon_{0}c_{0}\kappa f
\bm{g}^{-1}(\bm{\lambda}_{\kappa},\bm{\lambda}_{\kappa}) R_{\kappa}.
\non
\end{align}
\end{Proposition}
\begin{Proof}
They are derived from \fr{e-from-f}, \fr{h-from-f} and \fr{Z-kappa}. 
\qed
\end{Proof}
A consequence of Propositions\,\ref{fact-e-h-contact} and \ref{fact-e-h-by-R} 
is the following, which is
similar to Theorem 1.1 of \cite{Goto2024}. 
\begin{Corollary}[Existences of closed integral curves of $\wt{\bm{e}}^{\bm{g}}$ and $\wt{\bm{h}}^{\bm{g}}$]  
\label{fact-closed-field-lines}
Consider the case that $N$ is compact. Then there is at least one
closed integral curve of the vector field $\wt{\bm{e}}^{\bm{g}}$ on
$\{\{t\}\times N|\dot{f}(t)\neq 0\}\subset M$.
In addition, there is at least one
closed integral curve of the vector field $\wt{\bm{h}}^{\bm{g}}$
on $\{\{t\}\times N|f(t)\neq 0\}\subset M$.
\end{Corollary}  
\begin{Proof}
Recall the resolution of the Weinstein conjecture\,\cite{Hutchings2010},
that is, 
there is at least one closed orbit of the Reeb vector field on a closed  
contact manifold of dimension three.  Applying this resolution
with Propositions\,\ref{fact-e-h-contact} and \ref{fact-e-h-by-R}, 
one completes the proof.
\qed
\end{Proof}

In electromagnetism,  potentials for electromagnetic
fields are essential for theoretical developments.  
In the following, basic properties of such potentials are shown. 
Introduce the potential
$\bm{A}_{\kappa}\in \GamLamM{1}$ such that
$$
F=\dr \bm{A}_{\kappa},\qquad
\bm{A}_{\kappa}
=f\bm{\lambda}_{\kappa},
$$
where $f$ and $\bm{\lambda}_{\kappa}$
are given in Theorem\,\ref{fact-introduction-1}. 
The following is a consequence of Theorem\,\ref{fact-introduction-1}.
\begin{Corollary}[Invariance of $F$ and $\bm{A}_{\kappa}$] 
  \label{fact-contact-symplectic-maxwell}
The Maxwell field $F$ and its potential $\bm{A}_{\kappa}$ 
are invariant along $R_{\kappa}$:
$$
\cL_{R_{\kappa}}\bm{A}_{\kappa}
=0,\qquad\text{and}\qquad
\cL_{R_{\kappa}}F
=0.  
$$
\end{Corollary}
\begin{Proof}
It follows from the Cartan's formula that
\begin{align}
\cL_{R_{\kappa}}\bm{A}_{\kappa}
&=(\dr\ii_{R_{\kappa}}+\ii_{R_{\kappa}}\dr)\bm{A}_{\kappa}
\non\\
&=\dr f+\ii_{R_{\kappa}}(\dr f\wedge\bm{\lambda}_{\kappa}
+f\bdr\bm{\lambda}_{\kappa})
\non\\
&=\dr f-\dr f
\non\\
&=0.
\non
\end{align}
This also yields
$$
\cL_{R_{\kappa}}F
=\cL_{R_{\kappa}}\dr\bm{A}_{\kappa}
=\dr\cL_{R_{\kappa}}\bm{A}_{\kappa}
=0.
$$
\qed
\end{Proof}

\subsection{Proof of Theorem\,\ref{fact-introduction-2}}
In the following, Theorem\,\ref{fact-introduction-2}, in which 
the source $j_{4}$ is specified by \fr{j0} with
$$
\bm{J}=\varepsilon_{0}c_{0}f\sstar\bm{\lambda}_{\kappa},\qquad\text{and}\qquad
\rho=0,
$$
is proved by
modifying the last part of the proof of Theorem\,\ref{fact-introduction-1}. 

\Prooff{Proof of Theorem\,\ref{fact-introduction-2}}
Recall that $F=\dr (f\bm{\lambda}_{\kappa})$. 
Similar to the proof of Theorem\,\ref{fact-introduction-1},
it immediately follows that $\dr F=0$.
Similarly, it follows that
$$
\dr \star F
=-(c_{0}^{-2}\ddot{f}+\kappa^{2}f)(\sstar\bm{\lambda}_{\kappa})\wedge\dr x^{0}.
$$
Substituting $c_{0}^{-2}\ddot{f}+\kappa^{2}f=-f$ and
$j_{4}=\varepsilon_{0}f(\sstar\bm{\lambda}_{\kappa})\wedge\dr x^{0}$,
one has that
$$
\varepsilon_{0}\dr \star F
=\varepsilon_{0}f(\sstar\bm{\lambda}_{\kappa})\wedge\dr x^{0}
=j_{4}.
$$
Thus the theorem is proved.
\qed

The fields $\bm{e}$ and $\bm{h}$ can be extracted from $F$ as \fr{e-h-from-F}.
In addition, $\bm{D}$ and $\bm{B}$ can be obtained from
\fr{Maxwell-constitutive-relations-vacuum-decomposed}. They are 
\begin{align}
&\bm{e}
=\ii_{\partial/\partial x^{0}}F
=c_{0}^{-1}\dot{f}\bm{\lambda}_{\kappa},
\label{e-from-f-2}\\
&\bm{h}
=-\varepsilon_{0}c_{0}\ \ii_{\partial/\partial x^{0}}\star F
=-\varepsilon_{0}c_{0}\ \kappa f\bm{\lambda}_{\kappa},  
\label{h-from-f-2}\\
&\bm{D}
=\varepsilon_{0}\sstar\bm{e}
=\varepsilon_{0}c_{0}^{-1}\dot{f}\sstar\bm{\lambda}_{\kappa},
\non\\
&\bm{B}
=\mu_{0}\sstar\bm{h}
=-c_{0}^{-1}\kappa f\sstar\bm{\lambda}_{\kappa}.
\non
\end{align}
From these expressions, 
the Poynting two-form \fr{Poynting-2-form} vanishes identically,
$$
\bm{\cI}=0.
$$
In addition, it turns out that
the time-dependence of all the fields is written in terms of the 
sinusoidal functions, $\sin$ and $\cos$. 

Similar to Propositions\,\ref{fact-F-symplectic}, \ref{fact-e-h-contact} and
\ref{fact-e-h-by-R},  
$F$ in Theorem\,\ref{fact-introduction-2} has the following properties.
\begin{Proposition}[Maxwell form is symplectic 2] 
\label{fact-F-symplectic-London}
Let $F$ be the solution in Theorem\,\ref{fact-introduction-2}. 
When $f\dot{f}\neq 0$, this $F$ is symplectic.
\end{Proposition}
\begin{Proof}
A way to prove this is analogous to the proof for 
Proposition\,\ref{fact-F-symplectic}.
\qed
\end{Proof}
\begin{Proposition}[Electromagnetic one-forms are contact 2]
\label{fact-e-h-contact-London}
On the hypersurface $\{\{t\}\times N |\dot{f}(t)\neq0\}\subset M$,
the one-form $\bm{e}$ in \fr{e-from-f-2}is contact. In addition, 
on the hypersurface $\{\{t\}\times N| f(t)\neq0\}\subset M$, 
the one-form $\bm{h}$ in \fr{h-from-f-2} is contact.   
\end{Proposition}
\begin{Proof}
A way to prove this is analogous to the proof for 
Proposition\,\ref{fact-e-h-contact}.
\qed
\end{Proof}
\begin{Proposition}[Relations between the Reeb vector and electromagnetic fields 2]
\label{fact-e-h-by-R-London}
The vector fields $\wt{\bm{e}}^{\bm{g}}$ and $\wt{\bm{h}}^{\bm{g}}$ are written as
\begin{align}
\wt{\bm{e}}^{\bm{g}}
&=c_{0}^{-1}\dot{f}\bm{g}^{-1}(\bm{\lambda}_{\kappa},\bm{\lambda}_{\kappa})
R_{\kappa},
\non\\
\wt{\bm{h}}^{\bm{g}}
&=-\,\varepsilon_{0}c_{0}\kappa f
\bm{g}^{-1}(\bm{\lambda}_{\kappa},\bm{\lambda}_{\kappa}) R_{\kappa}.
\non
\end{align}
\end{Proposition}
\begin{Proof}
A way to prove this  is analogous to the proof for 
Proposition\,\ref{fact-e-h-by-R}. 
\qed
\end{Proof}

The energy densities \fr{energy-densities} are then written,
with emphasis on the $f$ dependence, as   
\begin{align}
\bm{\cE}_{\e}^{f}
&=\frac{1}{2}\bm{e}\wedge\bm{D}
=\frac{\varepsilon_{0}}{2}(c_{0}^{-1}\dot{f})^{2}
\bm{\lambda}_{\kappa}\wedge\sstar\bm{\lambda}_{\kappa},
\non\\
\bm{\cE}_{\m}^{f}
&=\frac{1}{2}\bm{h}\wedge\bm{B}
=\frac{\varepsilon_{0}}{2}(\kappa f)^{2}
\bm{\lambda}_{\kappa}\wedge\sstar\bm{\lambda}_{\kappa},
\non\\
\bm{\cE}_{\emem}^{f}
&=\bm{\cE}_{\e}+\bm{\cE}_{\m}
=\frac{\varepsilon_{0}}{2}\left\{
(c_{0}^{-1}\dot{f})^{2}+(\kappa f)^{2}
\right\}
\bm{\lambda}_{\kappa}\wedge\sstar\bm{\lambda}_{\kappa}.
\non
\end{align}
Then this $\bm{\cE}_{\emem}^{f}$ reduces as follows. 
Firstly, a general solution $f$ to \fr{ode2} can be expressed by 
$$
f(t)=f_{0}\cos(\omega_{\kappa}^{\prime}t),\qquad
\omega_{\kappa}^{\prime}
:=c_{0} \sqrt{\kappa^{2}+1},
$$
with $f_{0}$ being constant and by its time-shift $t\mapsto t+t_{0}$.    
Employing this expression, one has
$$
\left(c_{0}^{-1}\dot{f}\right)^{2}+\left(\kappa f\right)^{2}
= (\kappa^{2}+1)f_{0}^{2}\sin^{2}(\omega_{\kappa}^{\prime}t)
+\kappa^{2}f_{0}^{2}\cos^{2}(\omega_{\kappa}^{\prime}t),
$$
which yields 
$$
\bm{\cE}_{\emem}^{f}
=\frac{\varepsilon_{0}f_{0}^{2}}{2}
\left(\kappa^{2}+\sin^{2}(\omega_{\kappa}^{\prime}t)\right)
\bm{\lambda}_{\kappa}\wedge\sstar\bm{\lambda}_{\kappa}.
$$
Hence the integral of $\bm{\cE}_{\emem}^{f}$ over a non-empty region
$N^{\prime}\subset N$ depends on time.
Note that this time dependence of the integral does not
lead to an amplification of 
energy due to the oscillatory behavior in time.

\subsection{London equations associated with Theorem\,\ref{fact-introduction-2}}

From a broader perspective,
it is of interesting to see whether or not
the fields in Theorem\,\ref{fact-introduction-2}
are related to physical systems. 
In this subsection such relations are shown. 
In particular, the fields in Theorem\,\ref{fact-introduction-2}
is shown to be related to electromagnetic fields in superconductors.

The London equations describe external currents and vector potentials for
superconductors\,\cite{Sternberg1992}.
In the differential form language, it is written as the single equation
$$
\bm{j}=-c_{\text{Lon}}\bm{A},
$$
where $\bm{j}=\sstar\bm{J}$ with $c_{\text{Lon}}>0$ being constant, and
$\bm{A}$ is a potential one-form such that $\bm{B}=\bdr\bm{A}$. 
By applying the exterior derivative $\bdr$ to the both sides,
this equation leads to 
\beq
\bdr\bm{j}
=- c_{\text{Lon}}\bm{B}.
\label{j=-cB}
\eeq
In addition, it is known that the London equation leads to an equation
involving the Laplacian for 
the magnetic field. To write it in the differential form language, 
introduce the Laplacian $\bm{\Delta}$ acting on a form $\bm{\alpha}$
that does not contain $\dr x^{0}$, $\ii_{\partial/\partial x^{0}}\bm{\alpha}=0$:
$$
\bm{\Delta}\bm{\alpha}
:=(\sstar\bdr\sstar\bdr
+\bdr\sstar\bdr\sstar)\ \bm{\alpha}.
$$
It follows from this and $\sstar\sstar\bm{\alpha}=\bm{\alpha}$ that
$$
\sstar\bm{\Delta}\bm{\alpha}
=\bm{\Delta}\sstar\bm{\alpha}.
$$
Then the above mentioned equation for the magnetic field $\bm{B}$ associated
with the London equation is written as 
\beq
\bm{\Delta}\bm{B}
=c_{\kappa}\bm{B},
\label{Laplacian-B=c-B}
\eeq
with some constant $c_{\kappa}>0$. The quantity $c_{\kappa}^{-1/2}$ is called
the {\it London penetration depth} in the literature.
In addition to \fr{Laplacian-B=c-B},
there is a similar equation for $\bm{h}$. 

The following shows explicitly $c_{\text{Lon}}$ in \fr{j=-cB} 
and $c_{\kappa}$ in \fr{Laplacian-B=c-B} 
for the fields in Theorem\,\ref{fact-introduction-2}. 
\begin{Corollary}[London equations]
\label{fact-London-equations}
The fields in Theorem\,\ref{fact-introduction-2}  satisfy
$$
\bdr\bm{j}
=-c_{\text{Lon}}\bm{B},\qquad\text{where}\quad 
c_{\text{Lon}}
=\varepsilon_{0}c_{0}^{2}.
$$
In addition, it follows that 
$$
\bm{\Delta}\bm{B}
  =c_{\kappa}\bm{B},
  \quad\text{and}\quad
 \bm{\Delta}\bm{h}
=c_{\kappa}\bm{h},\qquad\text{where}\qquad  c_{\kappa}
=\kappa^{2}.
$$
\end{Corollary}    
\begin{Proof}
From 
\begin{align}
&\bm{j}=\sstar\bm{J}
=\varepsilon_{0}c_{0}f\bm{\lambda}_{\kappa},
\non\\
&\bm{B}=-c_{0}^{-1}f\kappa\sstar\bm{\lambda}_{\kappa},
\non
\end{align}
one has that
\begin{align}
\bdr\bm{j}
&=\varepsilon_{0}c_{0}f\bdr\bm{\lambda}_{\kappa}
\non\\
&=\varepsilon_{0}c_{0}f\kappa\sstar\bm{\lambda}_{\kappa}
\non\\
&=-\varepsilon_{0}c_{0}^{2}\bm{B}.
\non
\end{align}
From this, one has that $c_{\text{Lon}}=\varepsilon_{0}c_{0}^{2}$.  
In addition, it follows from
\begin{align}
\bdr\sstar\bdr\bm{\lambda}_{\kappa}
&=\kappa\bdr\bm{\lambda}
\non\\
&=\kappa^{2}\sstar\bm{\lambda}_{\kappa},
\non
\end{align}
and $\bdr\bm{B}=0$, \fr{Maxwell-decomposed-vacuum},  that  
\begin{align}
\bm{\Delta}\bm{B}
&=\bdr\sstar \bdr\sstar\bm{B}
\non\\
&=-c_{0}^{-1}\kappa f\bdr\sstar\bdr\bm{\lambda}_{\kappa}
\non\\
&=\kappa^{2}\bm{B}.
\non
\end{align}
With this, $\bm{h}=\mu_{0}^{-1}\sstar\bm{B}$, and 
$\sstar\bm{\Delta}=\bm{\Delta}\sstar$,
one has that
\begin{align}
\bm{\Delta}\bm{h}
&=\mu_{0}^{-1}\bm{\Delta}\sstar\bm{B}
\non\\
&=\mu_{0}^{-1}\sstar\bm{\Delta}\bm{B}
\non\\
&=\kappa^{2}\bm{h}.
\non
\end{align}
\qed
\end{Proof}
As shown in Corollary\,\ref{fact-London-equations}, 
a class of solutions to the two equations,
\fr{j=-cB} and \fr{Laplacian-B=c-B}, is constructed from the rotational 
Beltrami fields.
It is unclear to us whether another class of solutions exists
on Lorentzian manifolds (see \cite{Flores2026} for a case of G\"odel universe). 
From the perspective of microscopic physical models,
the following question arises: Can one link between a microscopic model of
superconductors and the present solution? The answer is not given
in this paper.  
Although it is beyond the scope of this paper,  
a detailed study on a microscopic model 
for an electron with a Beltrami field is expected to give a
microscopic interpretation of Corollary\,\ref{fact-London-equations}. 

\section{Liouville integrability and magnetic surfaces as level sets}
\label{section-magnetic-surfaces}
In this section magnetic surfaces are discussed by means of geometric methods.
Here magnetic surfaces are invariant two-dimensional surfaces
along magnetic vector fields in three-dimensional
spatial manifolds at fixed time $x^{0}$, and they have been discussed in
plasma and space sciences. After discussing the general theory,
some examples are provided.

\subsection{General theory}
Magnetic surfaces play key roles since
they form barriers and skeletons for magnetic fields.
Meanwhile, the rigorous definition of magnetic surface may be unclear
in the literature. 
To study magnetic surfaces quantitatively, in this paper,
its differential geometric definition is given as follows.
\begin{Definition}
\label{definition-magnetic-surface-physical}
Let $\wt{\bm{h}}^{\bm{g}}$ be a magnetic vector field which need not
be constructed from a Beltrami field.   
If there is a non-constant function $\Psi$ on $N$ 
such that
\beq
\cL_{\wt{\bm{h}^{\bm{g}}}}\Psi
=0,
\label{magnetic-surface-physical}
\eeq
then $\Psi^{-1}(c)\subset N$ is called a {\it magnetic surface}
associated with $\bm{h}$, where $c\in\mbbR$ is constant.
In addition $\Psi$ is called a {\it magnetic stream function} associated
with $\bm{h}$. 
\end{Definition}

Note that \fr{magnetic-surface-physical} is equivalent to
$\bdr\Psi(\wt{\bm{h}}^{\bm{g}})=0$, yielding
$$
\bm{g}\left(\wt{\bm{h}}^{\bm{g}},\wt{\bdr\Psi}^{\bm{g}}\right)
=0.
$$
This implies that $\wt{\bm{h}}^{\bm{g}}$ is orthogonal to 
$\wt{\bdr\Psi}^{\bm{g}}$. 
Hence, if there exists a $\Psi$ such that \fr{magnetic-surface-physical} holds,
then one can write it with an almost complex structure
${\mathcal J}^{\prime}:\GT{N}\to\GT{N}$ as 
\beq
\wt{\bm{h}}^{\bm{g}}
=\mathcal{J}^{\prime}\,\wt{\bdr\Psi}^{\bm{g}}.
\label{magnetic-stream-function-dual-h}
\eeq
With this $\mathcal{J}^{\prime}$, 
by defining ${\mathcal J}:\GamLam{N}{1}\to\GamLam{N}{1}$ 
appropriately, 
one can write the above equation as 
$$
\bm{h}
=\mathcal{J}\bdr\Psi.
$$
An analogue of \fr{magnetic-stream-function-dual-h} 
is known in hydrodynamics and is
of the form $v=\mathcal{J}^{\prime}\dr\Psi_{\text{fluid}}$, where 
$v$ expresses a velocity vector field of fluid 
and $\Psi_{\text{fluid}}$ is a stream function.  
Hence the term ``magnetic streaming function'' in
Definition\,\ref{definition-magnetic-surface-physical} is appropriate. 

In the case that $\wt{\bm{h}}^{\bm{g}}$ is proportional to the Reeb vector $R$, 
as in Proposition\, \ref{fact-e-h-by-R}, 
this $\Psi$ can be discussed in the contact geometric language as follows. 
\begin{Definition}
Let $(N,\ker\bm{\lambda})$ be a contact manifold,  
where $\bm{\lambda}$ need not be a rotational Beltrami one-form. 
If there is a non-constant function $\Psi$ on $N$ that is
conserved along the Reeb vector field $R$, 
$$
\cL_{R}\Psi
=0,
$$
then the function $\Psi$ is called a non-trivial
{\it first integral} of the Reeb vector field $R$.
\end{Definition}

One way to find a first integral is as follows.
\begin{Proposition}[First integral from a vector]
\label{fact-how-to-find-first-integral}
Let  $(N,\ker\bm{\lambda})$ be a contact manifold, 
where $\bm{\lambda}$ need not be a rotational Beltrami one-form.   
If a vector field $U$ satisfies
$$
\cL_{U}\bm{\lambda}
=0,
$$
and the function
$$
\Psi=\ii_{U}\bm{\lambda},
$$
is not constant, then $\Psi$ is a non-trivial first integral of $R$,
$\cL_{R}\Psi=0$. 
\end{Proposition}
\begin{Proof}
Applying the Cartan formula to $\cL_{U}\bm{\lambda}=0$, one has
$\ii_{U}\bdr\bm{\lambda}+\bdr\ii_{U}\bm{\lambda}=0$. Substituting 
$\Psi=\ii_{U}\bm{\lambda}$ to it, one has that
$$
\bdr\Psi
=-\ii_{U}\bdr\bm{\lambda}.
$$
Using the obtained equation above, one calculates
\begin{align}
\cL_{R}\Psi
&=\ii_{R}\bdr\Psi
\non\\
&=\ii_{R}(-\ii_{U}\bdr\bm{\lambda})
\non\\
&=\ii_{U}\ii_{R}\bdr\bm{\lambda}
\non\\
&=0.
\non
\end{align}
\qed
\end{Proof}
In Proposition\,\ref{fact-how-to-find-first-integral},
choosing $U$ as the Reeb vector field,
$U=R\not\in\ker\bm{\lambda}$, one has that
$\Psi=\ii_{R}\bm{\lambda}=1$.
This constant function is a 
trivial first integral, that is not a non-trivial first integral. 

Given a non-trivial first integral $\Psi$ of $R$, one then 
has the following vector field. 
\begin{Definition}
\label{Definition-horizontal-contact-Hamilton-vector}
Let $(N,\ker\bm{\lambda})$ be a contact manifold,
where $\bm{\lambda}$ need not be a rotational Beltrami one-form.
Assume that there exists a non-trivial first integral $\Psi$ of $R$.  
If a vector field $W_{\text{hor}}$ satisfies 
\beq
\ii_{W_{\text{hor}}}\bdr\bm{\lambda}
=-\bdr\Psi,\qquad\text{and}\qquad
\bm{\lambda}(W_{\text{hor}})
=0,
\label{horizontal-contact-Hamiltonian-vector-conditions}
\eeq
then $W_{\text{hor}}$ is called a
{\it horizontal contact Hamiltonian vector field} in this paper. 
\end{Definition}

In Definition\,\ref{Definition-horizontal-contact-Hamilton-vector},
$\Psi$ is a non-trivial first integral of $R$, 
and this first integral plays the role of a Hamiltonian
in $\ker\bm{\lambda}$.
As shown below, if $\Psi$ is not a first integral of $R$, then
there is an inconsistency. 
\begin{Remark}
Let $W_{h}$ be the Hamiltonian vector field associated with a function $h$
on $N$: 
\begin{align}
\ii_{W_{h}}\bdr\bm{\lambda}
=-\bdr h,\qquad\text{and}\qquad
\bm{\lambda}(W_{h})
=0.
\label{Hamiltonian-is-first-function-1}
\end{align}
By applying $\ii_{R}$ to the left hand side of the first equation of 
\fr{Hamiltonian-is-first-function-1}, one has that
$$
\ii_{R}\ii_{W_{h}}\bdr\bm{\lambda}
=-\ii_{W_{h}}\ii_{R}\bdr\bm{\lambda}
=0.
$$
By applying $\ii_{R}$ to the right hand side of the first equation of 
\fr{Hamiltonian-is-first-function-1}, one has that
$$
\ii_{R}(-\bdr h)
=-Rh.
$$
Since they are equal, one concludes that
$$
\cL_{R}h
=Rh=0.
$$
\end{Remark}

An interpretation of $W_{\text{hor}}$
in the physics language is given as follows.
Consider the so-called MHD (magneto-hydro-dynamics)
equation with a unit mass on a Riemannian manifold
$(N,\bm{g})$ with a connection $\nabla$:  
$$
\cL_{\partial/\partial t}\bm{v}+\nabla_{\wt{\bm{v}}^{\bm{g}}}\bm{v}
=\sstar\left(\bm{j}_{0}\wedge\bm{h}\right)-\bdr \Psi_{0}.
$$
where $\bm{v}\in\GamLam{N}{1}$ is a velocity of electrons,  
$\Psi_{0}\in\GamLam{N}{0}$ a pressure function,
and $\bm{j}_{0}\in\GamLam{N}{1}$ a current.
A {\it plasma equilibrium state} is such that
the left hand side equals zero.
A first step to understand dynamical systems is in general
to analyze an equilibrium state. 
To analyze such an equilibrium state,
a relation between $\bm{j}_{0}$ and $\bm{h}$ is necessary. 
This relation is $\bm{j}_{0}=\sstar\bdr\bm{h}$, which
is valid when a displacement current is negligible $\dot{\bm D}=0$, due to 
\fr{Maxwell-decomposed-vacuum}. 
These for equilibrium states yield that 
the system of equations for the electromagnetic force and the current: 
\beq
\sstar\left(\bm{j}_{0}\wedge\bm{h}\right)
=\bdr \Psi_{0},\qquad\text{and}\qquad
\bm{j}_{0}
=\sstar\bdr\bm{h}.
\label{plasma-equilibrium-full}
\eeq
Equation \fr{plasma-equilibrium-full} 
can be the point of departure for discussing 
magneto-hydrostatics\,\cite{Hudon2007,Perrella2023JMP}. 
Suppose that $\bm{h}$ is a rotational Beltrami one-form,
$\sstar\bdr\bm{h}=\kappa\bm{h}$ with $\kappa\neq0$ being constant. 
In this case, from  \fr{plasma-equilibrium-full}
and $\sstar\bm{h}=\kappa^{-1}\bdr\bm{h}$, it follows that
\beq
\ii_{\wt{\bm{j}_{0}}^{\bm{g}}}\bdr\bm{h}
=-\bdr(\kappa\Psi_{0}).
\label{plasma-equilibrium-1}
\eeq
Consider the component $\bm{j}_{0\,\text{hor}}$ of $\bm{j}_{0}$, such that
$\bm{h}$ and $\bm{j}_{0\,\text{hor}}$
are orthogonal each other with respect to $\bm{g}$:
$$
\bm{g}^{-1}\left(\bm{j}_{0\,\text{hor}},\bm{h}\right)
=0.
$$
which is written as the pairing: 
\beq
\bm{h}\left(\wt{\bm{j}_{0\,\text{hor}}}^{\bm{g}}\right)
=0.
\label{plasma-equilibrium-2}
\eeq 
The vector $\wt{\bm{j}_{0\,\text{hor}}}^{\bm{g}}$ is called a
{\it diamagnetic current}. 
To interpret $W_{\text{hor}}$ of 
Definition\,\ref{Definition-horizontal-contact-Hamilton-vector} physically,
compare \fr{horizontal-contact-Hamiltonian-vector-conditions}
and \fr{plasma-equilibrium-1} with \fr{plasma-equilibrium-2}. 
Then one has the correspondence list (see Table\,\ref{table-MHD-language}).

\begin{table}[htb]
\caption{Interpretations of horizontal contact Hamiltonian vector field}
\label{table-MHD-language}
\begin{center}
  \begin{tabular}{|c|c||c|c|}
    \hline
    &
    Geometry &
    &
    MHD
    \\ \hline
    $\bm{\lambda}$ &
    contact form &
    $\bm{h}$ &
    magnetic field
    \\ \hline 
    $W_{\text{hor}}$  &
    horizontal contact Hamiltonian vector field &
    $\bm{g}^{-1}(\bm{j}_{0\,\text{hor}},-)$ &
    diamagnetic current 
    \\ \hline
    $\Psi$ &
    first integral of the Reeb vector field &
    $\kappa\Psi_{0}$ &
    pressure 
    \\ \hline
  \end{tabular}
\end{center}    
\end{table}  

Similar to the case of Hamiltonian vector fields on symplectic manifolds,  
every horizontal contact vector field $W_{\text{hor}}$
is uniquely determined for a given first integral of $R$.   
To verify this uniqueness,  
recall that $\bdr\bm{\lambda}$ is symplectic in $\ker\bm{\lambda}$. By
definition every symplectic form is non-degenerate, this means that
$\bdr\bm{\lambda}$ is an isomorphism between a one-form and
a vector field.  
Hence the one-form $\bdr\Psi$ uniquely determines
$W_{\text{hor}}$ such that $\ii_{W_{\text{hor}}}\bdr\bm{\lambda}=-\bdr\Psi$ holds.
The following is an analogue of the conservation law
of energy in the symplectic case. 
\begin{Lemma}
\label{fact-conservation-law-horizontal-contact}
On $(N,\ker\bm{\lambda})$, a non-trivial first integral of the Reeb vector
field is conserved along
$W_{\text{hor}}$ that satisfying \fr{horizontal-contact-Hamiltonian-vector-conditions}:
$$
\cL_{W_{\text{hor}}}\Psi
=0.
$$
\end{Lemma}
\begin{Proof}
It follows from \fr{horizontal-contact-Hamiltonian-vector-conditions} that
\begin{align}
\cL_{W_{\text{hor}}}\Psi
&=\ii_{W_{\text{hor}}}\bdr\Psi
\non\\
&=\ii_{W_{\text{hor}}}(-\ii_{W_{\text{hor}}}\bdr\bm{\lambda})
\non\\
&=0.
\non
\end{align}
\qed
\end{Proof}
In addition to Lemma\,\ref{fact-conservation-law-horizontal-contact}, 
the following shows basic properties of $W_{\text{hor}}$. 
\begin{Proposition}[Commutator relation with $R$ and $W_{\text{hor}}$] 
On $(N,\ker\bm{\lambda})$, let $\Psi$ be a first integral of $R$,
and  $W_{\text{hor}}$  be the horizontal contact Hamiltonian vector field. 
Then $R$ and $W_{\text{hor}}$ satisfy
\begin{align}
\text{(i)} 
&\quad [R,W_{\text{hor}}]\in\ker{\bm\lambda},
\non\\
\text{(ii)}  
&\quad [R,W_{\text{hor}}]
=0.
\label{R-horizontal-contact-Hamiltonian-commute}
\end{align}
\end{Proposition}
\begin{Proof}
To prove (i), one starts with the
identity: 
$$
\cL_{R}\ii_{W_{\text{hor}}}\bm{\lambda}
-\ii_{W_{\text{hor}}}\cL_{R}\bm{\lambda}
=\ii_{[R,W_{\text{hor}}]}\bm{\lambda}.
$$
Since
$$\ii_{W_{\text{hor}}}\bm{\lambda}
=0,\qquad\text{and}\qquad 
\cL_{R}\bm{\lambda}
=(\ii_{R}\bdr+\bdr\ii_{R})\bm{\lambda}=0,
$$
one has $0=\ii_{[R,W_{\text{hor}}]}\bm{\lambda}$.
This means that $[R,W_{\text{hor}}]\in\ker\bm{\lambda}$. 

Our strategy to prove (ii) is as follows. 
The first step to show $[R,W_{\text{hor}}]=0$ is to derive    
$[R,W_{\text{hor}}]\in\ker\bdr\bm{\lambda}$. 
To prove $[R,W_{\text{hor}}]\in\ker\bdr\bm{\lambda}$, one uses the identity:
$$
\cL_{R}\ii_{W_{\text{hor}}}\bdr\bm{\lambda}
-\ii_{W_{\text{hor}}}\cL_{R}\bdr\bm{\lambda}
=\ii_{[R,W_{\text{hor}}]}\bdr\bm{\lambda}.
$$
Since
$$
\cL_{R}\ii_{W_{\text{hor}}}\bdr\bm{\lambda}=\cL_{R}(-\bdr\Psi)=-R\Psi=0
\qquad\text{and}\qquad 
\cL_{R}\bdr\bm{\lambda}=0,
$$
one has $0=\ii_{[R,W_{\text{hor}}]}\bdr\bm{\lambda}$.
Due to the non-degeneracy of
$\bdr\bm{\lambda}$ in $\ker\bm{\lambda}$, 
it follows that $[R,W_{\text{hor}}]=0$. 
\qed
\end{Proof}

Equation \fr{R-horizontal-contact-Hamiltonian-commute}
is a consequence from the definition of horizontal contact Hamiltonian vector
field.
Similar to this construction, 
an analogous commutator relation was derived in \cite{Burby2021}, 
and that analogous commutator relation is the main body of the definition
of integrable presymplectic system. 

An example of $\Psi$ with $W_{\text{hor}}$ is as follows.
\begin{Example}
Let $(\mbbR^{3},\ker\bm{\lambda})$ be a contact manifold with
with $\bm{\lambda}=\bdr x^{3}-x^{2}\bdr x^{1}$. The Reeb vector field is 
$$
R=\frac{\partial}{\partial x^{3}},
$$
and then any non-constant function $\Psi$ of $x^{1}$ and $x^{2}$,
$$
\Psi=\Psi(x^{1},x^{2}),
$$
is conserved along $R$, $\cL_{R}\Psi=0$. Hence this $\Psi$ is a non-trivial
first integral. 
Let $W_{\text{hor}}\in\ker\bm{\lambda}$ be written as
$$
W_{\text{hor}}
=a_{1}W_{1}+a_{2}W_{2},\qquad
W_{1}:=\frac{\partial}{\partial x^{2}},\qquad
W_{2}:=x^{2}\frac{\partial}{\partial x^{3}}+\frac{\partial}{\partial x^{1}},
$$
where $a_{1}$ and $a_{2}$ are functions  of $x^{1}$ and $x^{2}$, and 
they are uniquely determined from
\fr{horizontal-contact-Hamiltonian-vector-conditions} as
$$
a_{1}=\frac{\partial\Psi}{\partial x^{1}},\qquad\text{and}\qquad
a_{2}=-\frac{\partial\Psi}{\partial x^{2}}.
$$
This $\Psi$ is conserved along $W_{\text{hor}}$:
\begin{align}
W_{\text{hor}}\Psi
&=\frac{\partial\Psi}{\partial x^{1}}\frac{\partial}{\partial x^{2}}\Psi
-\frac{\partial\Psi}{\partial x^{2}}
\left(x^{2}\frac{\partial}{\partial x^{3}}+\frac{\partial}{\partial x^{1}}\right)\Psi
\non\\
&=0.
\non
\end{align}
With the standard metric on $N=\mbbR^{3}$,  
\beq
\bm{g}_{\mbbR^{3}}
=\dr x^{1}\otimes \dr x^{1}
+\dr x^{2}\otimes \dr x^{2}
+\dr x^{3}\otimes \dr x^{3}.
\label{g-E^3}
\eeq
the inner product between 
$W_{\text{hor}}$ and $R_{\kappa}$ is calculated to be 
\begin{align}
\bm{g}_{\mbbR^{3}}(-,R)
&=\bdr x^{3},
\non\\
\bm{g}_{\mbbR^{3}}(W_{\text{hor}},R)
&=-x^{2}\frac{\partial\Psi}{\partial x^{2}}.
\non
\end{align}
\end{Example}

When a non-trivial first integral $\Psi$ of the Reeb vector field has a special
property, more information about the horizontal contact Hamiltonian
vector field is available. To discuss this, recall that
a function $\Psi$ and a connection $\nabla$ on $N$
induce the {\it Hessian}, defined by 
\beq
(\text{Hess}^{\nabla}\Psi)(X,Y)
:=(\nabla_{X}\bdr \Psi)(Y),\qquad \forall X,Y\in\GT{N}.
\label{Hessian-XY-nabla-0}
\eeq
As it is shown below, at every point of the critical (point) set of $\Psi$,
$$
\Sigma_{\Psi}
:=\{\ p\in N\ |\ \bdr \Psi=0\ \},
$$
the Hessian $(\text{Hess}^{\nabla}\Psi)$ is independent of the 
choice of $\nabla$.  
The following Lemma can be applied to Riemannian manifolds of any dimension.
\begin{Lemma}
\label{fact-Hessian-at-critical-point}
Let $(N,\bm{g})$ be a Riemannian manifold,
and $\nabla$ be a connection on $N$. 
The Hessian of $\Psi$ is written as 
\beq
(\text{Hess}^{\nabla}\Psi)(X,Y)
=XY\Psi -(\bdr\Psi)(\nabla_{X}Y).
\label{Hessian-XY-nabla-2}
\eeq
Furthermore, at any point of $\Sigma_{\Psi}$,
it follows that 
\begin{align}
(\text{Hess}^{\nabla}\Psi)(X,Y)
=\ii_{X}\cL_{Y}\bdr\Psi.
\label{Hessian-XY-nabla-3}
\end{align}
\end{Lemma}
\begin{Proof}
Firstly, to rewrite \fr{Hessian-XY-nabla-0}, observe that
\begin{align}
XY\Psi
&=\nabla_{X}[(\bdr\Psi)(Y)]
\non\\
&=(\nabla_{X}\bdr\Psi)(Y)+(\bdr\Psi)(\nabla_{X}Y).
\non
\end{align}
Substituting this into \fr{Hessian-XY-nabla-0}, one has 
\fr{Hessian-XY-nabla-2}. 
Secondly, 
$(\text{Hess}^{\nabla}\Psi)(X,Y)-\ii_{X}\cL_{Y}\bdr\Psi$
is evaluated. Using the identity 
\begin{align}
\ii_{X}\cL_{Y}\bdr\Psi
&=\cL_{Y}\ii_{X}\bdr\Psi-\ii_{[Y,X]}\bdr\Psi
\non\\
&=YX\Psi-\bdr\Psi([Y,X]),
\non
\end{align}
and \fr{Hessian-XY-nabla-2}, one has
\begin{align}
(\text{Hess}^{\nabla}\Psi)(X,Y)-\ii_{X}\cL_{Y}\bdr\Psi
&=[X,Y]\Psi
-(\bdr\Psi)(\nabla_{X}Y)+\bdr \Psi([Y,X])
\non\\
&=-(\bdr\Psi)(\nabla_{X}Y).
\non
\end{align}
Thus, 
$$
(\text{Hess}^{\nabla}\Psi)(X,Y)-\ii_{X}\cL_{Y}\bdr\Psi
=-(\bdr\Psi)(\nabla_{X}Y).
$$
At any point of $\Sigma_{\Psi}$, one has
\fr{Hessian-XY-nabla-3}. 
\qed
\end{Proof}
From Lemma\,\ref{fact-Hessian-at-critical-point},
at a critical point $p\in\Sigma_{\Psi}$ the Hessian can be
written without $\nabla$ as $(\text{Hess}_{p}\Psi)$, 
and in the coordinates $(x^{1},x^{2},x^{3})$ of a three-dimensional manifold 
it is written as   
\begin{align}
&(\text{Hess}_{p}\Psi)
\left(\frac{\partial}{\partial x^{a}},\frac{\partial}{\partial x^{b}}
\right)
=\frac{\partial^{2}\Psi}{\partial x^{a}\partial x^{b}},
\non\\
&(\text{Hess}_{p}\Psi)\left(-,\frac{\partial}{\partial x^{a}}
\right)
=\sum_{i=1}^{3}\frac{\partial^{2}\Psi}{\partial x^{i}\partial x^{a}}\bdr x^{i}
=(\text{Hess}_{p}\Psi)\left(\frac{\partial}{\partial x^{a}},-\right),
\non\\
&(\text{Hess}_{p}\Psi)\left(-,-\right)
=\sum_{i=1}^{3}\sum_{j=1}^{3}
\frac{\partial^{2}\Psi}{\partial x^{i}\partial x^{j}}\bdr x^{i}\otimes\bdr x^{j},
\non
\end{align}
for all $a,b=1,2,3$.

The following is a basic property of the Hessian at critical points.
\begin{Lemma}
\label{fact-kernel-dimension-Hessian-critical-set}
Let $(N,\ker\bm{\lambda})$ be a contact manifold, where
$\bm{\lambda}$ need not be a rotational Beltrami one-form. 
If $\Psi$ is a first integral of $R$, then
$$
\dim(\ker(\text{Hess}\Psi))\geq 1\qquad \text{on}\quad\Sigma_{\Psi}.
$$
\end{Lemma}
\begin{Proof}
Applying $\cL_{Y}$ to the both sides of $\ii_{R}\dr\Psi=0$, one has
$$
\cL_{Y}\ii_{R}\bdr \Psi
=0,\qquad \forall Y\in\GT{N},
$$
which can be written using
$\cL_{Y}\ii_{R}\bdr\Psi-\ii_{R}\cL_{Y}\bdr\Psi=\ii_{[Y,R]}\bdr \Psi$ as
$$
\ii_{R}\cL_{Y}\bdr\Psi+\ii_{[Y,R]}\bdr\Psi
=0.
$$
Thus, $\bdr\Psi=0$ on $\Sigma_{\Psi}$, and \fr{Hessian-XY-nabla-3}
yield 
$$
(\text{Hess}\Psi)(R,Y)
=0,\qquad \text{on}\
\Sigma_{\Psi}.   
$$
This means that $R\in\ker(\text{Hess}\Psi)$ on $\Sigma_{\Psi}$.  
Combining this result with the fact that $R$ does not vanish, one has that
$\dim(\ker(\text{Hess}\Psi))\geq 1$.
\qed
\end{Proof}
Note that if $\Psi$ is constant on $N$ everywhere, a trivial first integral
of $R$, then
$\Sigma_{\Psi}=\{p\in N |\bdr\Psi=0\}=N$. In this case from 
$$
\ker(\text{Hess}_{p}\Psi)
=T_{p}\Sigma_{\Psi}
=T_{p}N, \qquad
\forall p\in N,
$$
and $\dim T_{p}N=\dim N$,  it follows that
$$
\dim\ker(\text{Hess}_{p}\Psi)
=\dim(T_{p}N)
=\dim N.\qquad \forall p\in N.
$$
This equation does not hold in the case that $\Psi$ is a
non-trivial first integral.
Meanwhile, in the case that $\Psi$ is a class of non-trivial first integrals,   
as shown below, another equality holds for the dimension of
$\ker(\text{Hess}_{p}\Psi)$.

The following is introduced to classify non-trivial first integrals of $R$. 
\begin{Definition}
A {\it Morse-Bott function} $\Psi$ on a manifold $N$ is a function 
\begin{enumerate}
\item
whose critical set $\Sigma_{\Psi}$
is a submanifold, and  
\item 
whose Hessian $(\text{Hess}_{p}\Psi)$, $\forall p\in\Sigma_{\Psi}$,   
is non-degenerate in transverse directions along 
$\Sigma_{\Psi}$: 
$$
\ker(\text{Hess}_{p}\Psi)
=T_{p}\Sigma_{\Psi},\qquad \forall p\in \Sigma_{\Psi}.
$$
\end{enumerate}
\end{Definition}
It immediately follows that
$$
\dim\ker(\text{Hess}_{p}\Psi)
=\dim(T_{p}\Sigma_{\Psi})
=\dim\Sigma_{\Psi},\qquad \forall p\in \Sigma_{\Psi}.
$$

As written in \cite{Geiges2024},
the Reeb flow is called {\it Bott integrable} if there is a
Morse-Bott function $\Psi$ that is invariant under the Reeb flow, and
$\Psi$ is called a {\it Bott integral} of $R$. 

\begin{Remark}
\label{fact-Morse-Bott-dimensions}
At a point $p\in\Sigma_{\Psi}$ with $\Psi$ being a
non-trivial first integral of $R$,
it follows by definition that $\bdr\Psi=0$.
From this, $\ii_{W_{\text{hor}}}\bdr\bm{\lambda}=-\bdr\Psi$,
and  the non-degenerate property of $\bdr\bm{\lambda}$ in $\ker\bm{\lambda}$,
it follows that
$$
W_{\text{hor}}|_{p}
=0,\qquad \forall p\in\Sigma_{\Psi}.
$$
This means that $p\in\Sigma_{\Psi}$ is a singular point of
the vector field $W_{\text{hor}}$.
This singular point is equivalent to a fixed point of the 
system for a curve, $\dot{\gamma}=W_{\text{hor}}|_{\gamma}$. 
The totality of the singular points is then 
written in this paper as 
$$
\text{Fix}(W_{\text{hor}})
=\{\ p\in N\ |\ W_{\text{hor}}|_{p}
=0 \ \}.
$$
Thus,
$$
\text{Fix}(W_{\text{hor}})
=\Sigma_{\Psi},
$$
and then 
the dimension of the singular point set is obtained
as
$$
\dim(\text{Fix}(W_{\text{hor}}))
=\dim(\Sigma_{\Psi}).
$$
\end{Remark}

To introduce the notion of integrability in electromagnetism,
we discuss the Liouville integrability on a manifold obtained by the 
symplectization of a contact manifold 
(see Section\,\ref{section-review-symplectic-contact}).  
Recall here that $\iota_{*}X$ is often abbreviated simply as $X$, where
$X$ is a vector field on $N$ and $\iota:N\to M$ is the embedding. 

\begin{Proposition}[Liouville integrability and first integrals]
\label{fact-first-Liouville-integrable}
Let $(N,\ker\bm{\lambda})$ be a three-dimensional
contact manifold, where
$\bm{\lambda}$ need not be a rotational Beltrami one-form, 
and $f_{0}$ be a function on $\mbbR$.   
Let $M=\mbbR\times N$ and $F=\dr (f_{0}\bm{\lambda})$, so that
$(M_{0}, F)$ is the symplectic manifold, that is, 
$M_{0}:=M\setminus\{\dr f_{0}\neq 0\}\cup\{f_{0}\neq 0\}$.   
Assume that a non-trivial first integral $\Psi$ of the Reeb vector field
exists.  Then
$(f_{0},\Psi)$ on $M_{0}$ is a Liouville integrable system. 
\end{Proposition}
\begin{Proof}
The symplectic form is given by 
$$
F=\dr f_{0}\wedge\bm{\lambda}+f_{0}\bdr\bm{\lambda}.
$$
This is indeed symplectic on $M_{0}$, because $\dr F=0$ and 
$$
F\wedge F
=2f_{0}\dr f_{0}\wedge\bm{\lambda}\wedge\bdr\bm{\lambda}
\neq 0,
$$
everywhere on $M_{0}$. Let
$$
X_{\text{hor}}
=\frac{1}{f_{0}}W_{\text{hor}}.
$$
Firstly, it is shown here that the Reeb vector $R$ and $X_{\text{hor}}$ above
are Hamiltonian vector fields on $M_{0}$.  
The interior product between $R$ and $F$ is calculated to yield
$R=X_{f_{0}}$: 
$$
\ii_{R}F
=-\dr f_{0}.
$$
The interior product between $X_{\text{hor}}$ and $F$ is calculated to
yield $X_{\text{hor}}=X_{\Psi}$: 
\begin{align}
\ii_{X_{\text{hor}}}F
&=\frac{1}{f_{0}}\ii_{W_{\text{hor}}}(\dr f_{0}\wedge\bm{\lambda}
+f_{0}\bdr\bm{\lambda})
\non\\
&=\ii_{W_{\text{hor}}}\bdr\bm{\lambda}
\non\\
&=-\bdr\Psi,
\non
\end{align}
where \fr{horizontal-contact-Hamiltonian-vector-conditions} has been
used in the last equality. 
Hence $R$ and $X_{\text{hor}}$ are Hamiltonian vector fields with $f_{0}$ and
$\Psi$, respectively.  

Secondly, the functions $f_{0}$ and $\Psi$ are shown to be involutive.
Since $f_{0}$ is a function on $\mbbR$ and $\Psi$ is a function on $N$, 
one has that 
$$
\{f_{0},\Psi\}_{F}
=F(X_{f_{0}},X_{\Psi})
=X_{\text{hor}}f_{0}
=0.
$$

Finally, the functionary independence is shown.
By recalling that $\Psi$ is not constant by assumption and
that $\dr f_{0}\neq 0$, it follows that 
$$
\dr f_{0}\wedge\bdr \Psi
\neq 0.
$$

Combining these, one concludes that
$(f_{0},\Psi)$ forms an integrable system on $(M_{0},F)$.  
\qed
\end{Proof}
In Proposition\,\ref{fact-first-Liouville-integrable},
the key assumption is the existence of a non-trivial first integral of
the Reeb vector field. When this integral exists and it is Morse-Bott,
the Reeb flow is called {\it Bott integrable}. 
In the case of three-dimensional manifolds,
the question of when a Reeb vector field is Bott integrable has been 
answered\,\cite{Geiges2024}.
To apply Proposition\,\ref{fact-first-Liouville-integrable} to Maxwell fields,
$f_{0}$ is chosen to be a solution to \fr{ode1}. 

Recall that, in the case that magnetic vector field is constructed from
a rotational Beltrami field, the Reeb vector field 
is proportional to a magnetic vector field
$\wt{\bm{h}}^{\bm{g}}=\bm{g}^{-1}(\bm{h},-)$.
This is stated in Propositions\,\ref{fact-e-h-by-R} and
\,\ref{fact-e-h-by-R-London} as 
\begin{align}
\wt{\bm{h}}^{\bm{g}}
&=-\varepsilon_{0}c_{0}\kappa f \bm{g}^{-1}(\bm{\lambda}_{\kappa},\bm{\lambda}_{\kappa})R_{\kappa}.
\non
\end{align}
Thus, 
$$
\cL_{\wt{\bm{h}}^{\bm{g}}}\Psi
=0
\qquad\Longleftrightarrow\qquad
\cL_{R_{\kappa}}\Psi
=0.
$$
In other words, non-trivial first integrals are equivalent to magnetic
stream functions (see Definition\,\ref{definition-magnetic-surface-physical}).

Applying Proposition\,\ref{fact-first-Liouville-integrable} to
Maxwell's equations, one has the following that links Maxwell's equations and
integrable Hamiltonian systems. 
\begin{Theorem}[Magnetic surface implies integrability]
\label{fact-magnetic-surface-integrability}
Let $\bm{h}$ be given in \fr{h-from-f} or \fr{h-from-f-2}.
If there is a magnetic stream function associated with $\bm{h}$, then
Maxwell's equations are a Liouville integrable system on $M$.  
\end{Theorem}
\begin{Proof}
Since there is a magnetic stream function, or equivalently
a non-trivial first integral $\Psi$, one can apply
Proposition\,\ref{fact-first-Liouville-integrable}.
\qed
\end{Proof}
    
\subsection{Examples of magnetic surfaces as integrals}
Several examples of magnetic surfaces are shown below.

\subsubsection{Yoshida's example}
Let $N=\mbbR^{3}$ with the coordinates
$(x^{1},x^{2},x^{3})=(x,y,z)$, and $\bm{\lambda}$ be 
$$
\bm{\lambda}
=\sin\phi(z)\bdr x+\cos\phi(z)\bdr y,
$$
where $\phi$ is an arbitrary function of $z$. 
This was introduced in \cite{YoshidaZ1991Butsuri}, and is a weak version of
Beltrami:
\begin{align}
\sstar\bdr\bm{\lambda}
&=\phi^{\prime}(z)[\cos\phi\bdr y+\sin\phi\bdr x]
\non\\
&=\phi^{\prime}(z)\bm{\lambda},
\non
\end{align}
where $\phi^{\prime}(z)=\dr\phi/\dr z$.  
The Reeb vector field is
$$
R=\sin\phi\frac{\partial}{\partial x}+\cos\phi\frac{\partial}{\partial y},
$$
which shows that any non-constant function of $z$,
$$
\Psi=\Psi(z),
$$
is a non-trivial first integral, $R\Psi(z)=0$.
Hence $\Psi^{-1}(c)\subset\mbbR^{3}$ forms a level set with $c\in\mbbR$. 
In the special case $\Psi(z)=z^{2}$, this function is Morse-Bott, that is, 
$$
T_{p}\Sigma_{\Psi}
=\ker(\text{Hess}_{p}\Psi)
=\text{span}\left\{\ 
\frac{\partial}{\partial x},\frac{\partial}{\partial y}\  
\right\}, \quad\forall p\in\Sigma_{\Psi},
$$
which is a two-dimensional vector space and is verified from  
$$
\Sigma_{\Psi}
=\{(x,y,z)\in\mbbR^{3}|z=0\}, 
\qquad\text{and}\qquad
\text{Hess}_{p}\Psi
=2\bdr z\otimes \bdr z,
\quad\forall p\in\Sigma_{\Psi}.
$$

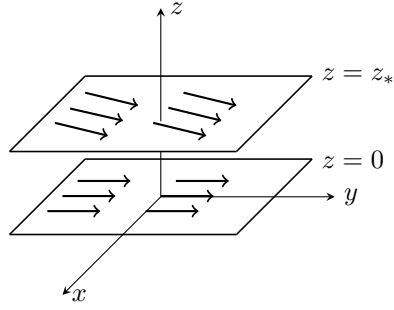
\begin{figure}[htbp]
\begin{center}
\begin{tikzpicture}[scale=1.0]
\draw[->,>=stealth] (0,0)--(-1.3,-1.3) node[right] {$x$}; 
\draw[->,>=stealth] (0,0)--(2.3,0) node[right] {$y$}; 
\draw (0,0)--(0,0.6) ; 
\draw[->,>=stealth] (0,1)--(0,2.5) node[right] {$z$}; 
\draw[semithick] (-2.0,-0.5)--(1.0,-0.5);
\draw[semithick] (-2.0,-0.5)--(-1.0,0.5);
\draw[semithick] (1.0,-0.5)--(2.0,0.5) node[right] {$z=0$};
\draw[semithick] (-1.0,0.5)--(2.0,0.5);
\draw (-3.8,1.5) node[right] {.};
\draw[semithick] (-2.0,0.6)--(1.0,0.6);
\draw[semithick] (-2.0,0.6)--(-1.0,1.6);
\draw[semithick] (1.0,0.6)--(2.0,1.6) node[right] {$z=z_{*}$};
\draw[semithick] (-1.0,1.6)--(2.0,1.6);
\draw[thick, ->](0,0.01)--(0.7,0.01) ;
\draw[thick, ->](0+0.2,0.01+0.2)--(0.7+0.2,0.01+0.2) ;
\draw[thick, ->](0-0.2,0.01-0.2)--(0.7-0.2,0.01-0.2) ;
\draw[thick, ->](0-1.3,0.01)--(0.7-1.3,0.01) ;
\draw[thick, ->](0-1.3+0.2,0.01+0.2)--(0.7-1.3+0.2,0.01+0.2) ;
\draw[thick, ->](0-1.3-0.2,0.01-0.2)--(0.7-1.3-0.2,0.01-0.2) ;
\draw[thick, ->](0.1,1.01+0.17)--(0.8,1.01) ;
\draw[thick, ->](0.1+0.2,1.01+0.2+0.17)--(0.8+0.2,1.01+0.2) ;
\draw[thick, ->](0.1-0.2,1.01-0.2+0.17)--(0.8-0.2,1.01-0.2) ;
\draw[thick, ->](0.1-1.3,1.01+0.16)--(0.8-1.3,1.01) ;
\draw[thick, ->](0.1-1.3+0.2,1.01+0.2+0.17)--(0.8-1.3+0.2,1.01+0.2) ;
\draw[thick, ->](0.1-1.3-0.2,1.01-0.2+0.17)--(0.8-1.3-0.2,1.01-0.2) ;
\end{tikzpicture}
\end{center}
\caption{ The Reeb vector fields on the level sets
  $\Psi^{-1}(0)$ and $\Psi^{-1}(z_{*})$
  for $\phi(z)=\kappa z$ and $\Psi(z)=z$, $\kappa\neq0$.
  On the plane $\{(x,y,0)|x,y\in\mbbR\}$ drawn at the bottom,
  the Reeb vector field is $R_{\kappa}=\partial/\partial y$.
  On the plane
  $\{(x,y,z_{*})|x,y\in\mbbR,\sin(\kappa z_{*})>0, \cos(\kappa z_{*})>0\}$ 
  drawn on the top, the Reeb vector field is
  $R_{\kappa}=\sin(\kappa z_{*})\partial/\partial x+\cos(\kappa z_{*})\partial/\partial y$.}
\label{figure-Reeb-plane}
\end{figure}

Choose $\phi(z)=\kappa z$ with $\kappa$ being a non-zero constant,
and write the contact form as
$\bm{\lambda}_{\kappa}$.
From straightforward calculations it follows that
$\sstar\bdr\bm{\lambda}_{\kappa}=\kappa\bm{\lambda}_{\kappa}$. 
Recall that the magnetic vector field is obtained from \fr{h-from-f} as 
$\wt{\bm{h}}^{\bm{g}}=-\varepsilon_{0}c_{0}\kappa f
\wt{\bm{\lambda}}_{\kappa}^{\bm{g}}$. 
Then non-constant arbitrary function $\Psi$ of $z$
is a magnetic stream function, and 
$\Psi^{-1}(c)$ is a magnetic surface, where this surface is a $(x,y)$-plane.  
Figure\,\ref{figure-Reeb-plane} shows 
the Reeb vector fields $R_{\kappa}$ on the level sets $\Psi^{-1}(c)\subset\mbbR^{3}$ with two different values of $c\in\mbbR$. 
For the first integral $\Psi$, the vector field satisfying
\fr{horizontal-contact-Hamiltonian-vector-conditions} is obtained as follows. 
Expand $W_{\text{hor}}$ in terms of the basis of $\ker\bm{\lambda}_{\kappa}$ as
$$
W_{\text{hor}}
=a_{1}\frac{\partial }{\partial z}
+a_{2}\left(\cos(\kappa z)\frac{\partial}{\partial x}-\sin(\kappa z)\frac{\partial}{\partial y}\right),
$$
with $a_{1}$ and $a_{2}$ functions to be determined.
From the condition that the two equations are equal:
\begin{align}
\ii_{W_{\text{hor}}}\bdr\bm{\lambda}
&=a_{1}\kappa(\cos(\kappa z)\bdr x-\sin(\kappa z)\bdr y)
-a_{2}\kappa\bdr z,
\non\\
-\bdr\Psi
&=-\Psi^{\prime}\bdr z,\qquad
\Psi^{\prime}(z)
:=\frac{\dr\Psi}{\dr z}(z),
\non
\end{align}
one obtains $a_{1}=0$ and $a_{2}=\Psi^{\prime}/\kappa$:
$$
W_{\text{hor}}
=\frac{\Psi^{\prime}}{\kappa}\left(
\cos(\kappa z)\frac{\partial}{\partial x}
-\sin(\kappa z)\frac{\partial}{\partial y}
\right).
$$
With the standard metric \fr{g-E^3}, 
and from
\begin{align}
\bm{g}_{\mbbR^{3}}(-,W_{\text{hor}})
&=\frac{\Psi^{\prime}(z)}{\kappa}(\cos(\kappa z)\bdr x-\sin(\kappa z)\bdr y),
\non\\
\bm{g}_{\mbbR^{3}}(R_{\kappa},W_{\text{hor}})
&=\frac{\Psi^{\prime}(z)}{\kappa}\left[\cos(\kappa z)(R_{\kappa}x)
-\sin(\kappa z)(R_{\kappa}y)\right],
\non
\end{align}
one obtains the orthogonal relation between 
$W_{\text{hor}}$ and $R_{\kappa}$: 
$$
\bm{g}_{\mbbR^{3}}(R_{\kappa},W_{\text{hor}})
=0.
$$
When $\Psi=z^{2}$, a Morse-Bott function, it follows that
$$
\dim(\text{Fix}(W_{\text{hor}}))
=\dim(\Sigma_{\Psi})
=2.
$$ 

There are similar contact one-forms. For example, see
Example II.14 of \cite{Perrella2023JMP} for the case of $\kappa=1$, and
Example\,\ref{example-Beltrami-Giroux} of this paper
on $N=S^{1}\times S^{1}\times S^{1}$. 

\subsubsection{The Lundquist model}

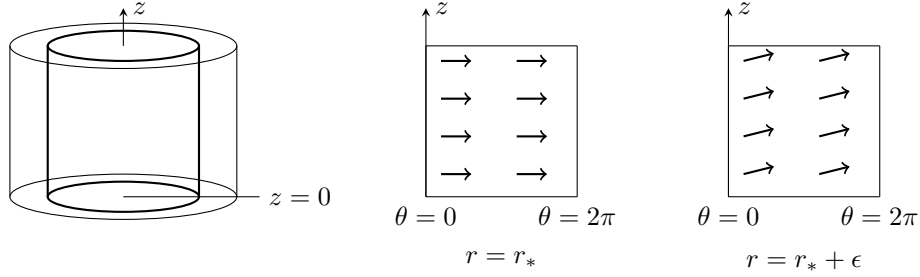
\begin{figure}[htbp]
  \begin{center}
\begin{tikzpicture}[scale=1.0]
%
\draw[->,>=stealth] (0,2.0)--(0,2.5) node[right] {$z$}; 
\draw (0,0)--(1.8,0) node[right] {$z=0$};
\draw (-1.5,0) -- (-1.5,2);
\draw (1.5,0) -- (1.5,2);
  
\draw (0,0) ellipse (1.5 and 0.3);
\draw (0,2) ellipse (1.5 and 0.3);
\draw[thick] (-1,0)--(-1,2);
\draw[thick] (1,0)--(1,2);
\draw[thick] (0,0) ellipse (1 and 0.2);
\draw[thick] (0,2) ellipse (1 and 0.2);
\draw[->,>=stealth] (4,0)--(4,2.5) node[right] {$z$}; 
\draw (4,2)--(4,0) node[below] {$\theta=0$};
\draw (6,2)--(6,0) node[below] {$\theta=2\pi$};
\draw (4,0)--(6,0);
\draw (4,2)--(6,2);
\node[below] at (5,-0.6) {$r=r_{*}$};
\foreach \x in{0.1, 0.6, 1.1, 1.6} \draw[thick,->] (4.2,0.2+\x)--(4.2+0.4,0.2+\x);
\foreach \x in{0.1, 0.6, 1.1, 1.6} \draw[thick,->] (5.2,0.2+\x)--(5.2+0.4,0.2+\x);
\draw[->,>=stealth] (8,0)--(8,2.5) node[right] {$z$}; 
\draw (8,2)--(8,0) node[below] {$\theta=0$};
\draw (10,2)--(10,0) node[below] {$\theta=2\pi$};
\draw (8,0)--(10,0);
\draw (8,2)--(10,2);
\node[below] at (9,-0.6) {$r=r_{*}+\epsilon$};
\foreach \x in{0.1, 0.6, 1.1, 1.6} \draw[thick,->] (8.2,0.2+\x)--(8.2+0.4,0.2+\x+0.1);
\foreach \x in{0.1, 0.6, 1.1, 1.6} \draw[thick,->] (9.2,0.2+\x)--(9.2+0.4,0.2+\x+0.1);

\end{tikzpicture}
\end{center}
\caption{(Left) Two level sets 
  $\Psi^{-1}(c)\subset\mbbR^{3}$ with two different
  values of $c\in\mbbR$ for $\Psi(r)=r$, drawn with thick lines and
  with the standard lines. 
  (Middle) The unfolded diagram of the level
  set $\Psi^{-1}(r_{*})\subset\mbbR^{3}$, where
  $J_{0}(\kappa r_{*})=0$ is satisfied. The Reeb vector field is parallel
  to $\partial/\partial \theta$,
  $R_{\kappa}=[r_{*}J_{1}(\kappa r_{*})]^{-1}\partial/\partial \theta$,
  from which the Reeb orbits are closed. 
  (Right) The unfolded diagram of the level set
  $\Psi^{-1}(r_{*}+\epsilon)\subset\mbbR^{3}$, 
  where $\epsilon>0$ is a constant such that  
  $J_{0}(\kappa (r_{*}+\epsilon))>0$ and $J_{1}(\kappa (r_{*}+\epsilon))>0$
  are satisfied. 
  The Reeb vector field is not parallel to $\partial/\partial\theta$,
  from which the Reeb orbits are not closed on this level set.
}
\label{figure-Reeb-Lundquist}
\end{figure}

Let $N=\mbbR^{3}$ with the cylindrical coordinates $(r,\theta,z)$ ,
and $\bm{\lambda}_{\kappa}$ be 
\beq
\bm{\lambda}_{\kappa}
= J_{0}(\kappa r)\bdr z+J_{1}(\kappa r)r\bdr\theta, 
\label{Lundquist-form-b0}
\eeq
where  $\kappa>0$ is a constant, and  
$J_{0}$ and $J_{1}$ are the zero-th and first Bessel functions.
The magnetic vector field with \fr{h-from-f} 
is often discussed as a simple model of magnetic fields\,\cite{Vandas2017}.
In this paper $\bm{\lambda}_{\kappa}$ in \fr{Lundquist-form-b0}
is called the {\it Lundquist form}. 
This is a rotational Beltrami one-form:
$$
\sstar\bdr\bm{\lambda}_{\kappa}
=\kappa \bm{\lambda}_{\kappa}.
$$
Simple calculations show that the Reeb vector associated with
$\bm{\lambda}_{\kappa}$ is
$$
R_{\kappa}
=\frac{1}{J_{0}^{2}(\kappa r)+J_{1}^{2}(\kappa r)}
\left[
J_{1}(\kappa r)\frac{1}{r}\frac{\partial}{\partial \theta}
+  J_{0}(\kappa r)\frac{\partial}{\partial z}
\right].
$$
Note that at $r_{*}$ satisfying $J_{0}(\kappa r_{*})=0$, the Reeb orbit is
a closed curve on a $z=\const$ plane.
This closed curve at $r=r_{*}$ is obtained by solving the system of
the differential equations 
$$
\frac{\dr\theta}{\dr s}
=\frac{1}{r_{*}J_{1}(\kappa r_{*})},\quad\text{and}\quad
\frac{\dr z}{\dr s}
=0  
$$
for $\theta$ and $z$ of $s\in\mbbR$. 
In addition, it is obvious that any non-constant
function of $r$,
$$
\Psi=\Psi(r),
$$
is a non-trivial first integral of $R_{\kappa}$, $\cL_{R_{\kappa}}\Psi=0$. 
Figure\,\ref{figure-Reeb-Lundquist} shows 
the Reeb vector fields on the level sets
$\Psi^{-1}(c)\subset\mbbR^{3}$ with two different values of $c\in\mbbR$. 

As a special case, let 
$$
\Psi_{0}(r,\theta,z)
=(r-r_{0})^{2},
$$
with $r_{0}\geq 0$ being a fixed constant.
As shown below, the function $\Psi_{0}$ is Morse-Bott.
\begin{enumerate}
\item
When $r_{0}=0$, by introducing the Cartesian coordinates $(x,y,z)$,
it follows that   
$$
\Sigma_{\Psi_{0}}
=\{(x,y,z)|x=y=0\},\quad\text{and}\quad
\text{Hess}_{p}\Psi_{0}
=2\bdr x\otimes \bdr x +2\bdr y\otimes \bdr y,
\qquad \forall p\in\Sigma_{\Psi_{0}}.
$$
Thus, 
$$
T_{p}\Sigma_{\Psi_{0}}
=\ker(\text{Hess}_{p}\Psi_{0})
=\text{span}\left\{\ \frac{\partial}{\partial z}\ \right\},
\qquad \forall p\in\Sigma_{\Psi_{0}}. 
$$

\item
When $r_{0}> 0$, it follows that   
$$
\Sigma_{\Psi_{0}}
=\{(r,\theta,z)|r=r_{0}\},\qquad\text{and}\qquad
\text{Hess}_{p}\Psi_{0}
=2\bdr r\otimes \bdr r,
\qquad \forall p\in\Sigma_{\Psi_{0}}. 
$$
Thus, 
$$
T_{p}\Sigma_{\Psi_{0}}
=\ker(\text{Hess}_{p}\Psi_{0})
=\text{span}\left\{\ \frac{\partial}{\partial \theta},\frac{\partial}{\partial z}\ \right\},
\qquad \forall p\in\Sigma_{\Psi_{0}}. 
$$

\end{enumerate}

To obtain the horizontal contact Hamiltonian vector field,
decompose $W_{\text{hor}}$ in terms of the
basis of $\ker\bm{\lambda}_{\kappa}$ as 
$$
W_{\text{hor}}
=a_{1}\frac{\partial}{\partial r}+a_{2}
\left(J_{0}(\kappa r)\frac{1}{r}\frac{\partial}{\partial \theta}
-J_{1}(\kappa r)\frac{\partial}{\partial z}
\right),
$$
with $a_{1}$ and $a_{2}$ some functions to be determined.
Straightforward calculations show that 
\begin{align}
\bdr\bm{\lambda}_{\kappa}
&=-\kappa J_{1}(\kappa r)\bdr r\wedge\bdr z
+\kappa J_{0}(\kappa r)\bdr r\wedge r\bdr \theta 
\non\\
\ii_{W_{\text{hor}}}\bdr\bm{\lambda}_{\kappa}
&=-\kappa J_{1}(\kappa r)\left[a_{1}\bdr z+a_{2} J_{1}(\kappa r)\bdr r\right]
+\kappa J_{0}(\kappa r)
\left[a_{1}r\bdr \theta -a_{2}J_{0}(\kappa r)\bdr r\right]
\non\\
&=-\kappa J_{1}(\kappa r) a_{1}\bdr z+\kappa J_{0}(\kappa r) a_{1}r\bdr \theta
-\kappa\left[J_{0}^{2}(\kappa r)+J_{1}^{2}(\kappa r)\right]a_{2}\bdr r,
\non
\end{align}
and $\bdr\Psi=(\dr\Psi/\dr r)\bdr r$.
Substituting these into \fr{horizontal-contact-Hamiltonian-vector-conditions},
one obtains $a_{1}$ and $a_{2}$, yielding 
$$
W_{\text{hor}}
=\frac{1}{\kappa (J_{0}^{2}(\kappa r)+J_{1}^{2}(\kappa r))}\frac{\dr\Psi}{\dr r}
\left(J_{0}(\kappa r)\frac{1}{r}\frac{\partial}{\partial \theta}
-J_{1}(\kappa r)\frac{\partial}{\partial z}
\right).
$$
With the metric from \fr{g-E^3}, 
$$
\bm{g}
=\bdr r\otimes \bdr r+r^{2}\bdr\theta\otimes\bdr\theta
+\bdr z\otimes \bdr z, 
$$
and from
\begin{align}
\bm{g}(-,W_{\text{hor}})
&=\frac{1}{\kappa (J_{0}^{2}(\kappa r)+J_{1}^{2}(\kappa r))}
\frac{\dr\Psi}{\dr r}(J_{0}(\kappa r)r\bdr \theta-J_{1}(\kappa r)\bdr z),
\non\\
\bm{g}(R,W_{\text{hor}})
&=\frac{1}{\kappa (J_{0}^{2}(\kappa r)+J_{1}^{2}(\kappa r))^{2}}
\frac{\dr\Psi}{\dr r}(J_{0}(\kappa r)J_{1}(\kappa r) -J_{1}(\kappa r)J_{0}(\kappa r)),
\non
\end{align}
one obtains the orthogonal relation between $R_{\kappa}$ and $W_{\text{hor}}$:
$$
\bm{g}(R_{\kappa},W_{\text{hor}})
=0.
$$
When $\Psi=\Psi_{0}$, a Morse-Bott function, it follows that
$$
\dim(\text{Fix}(W_{\text{hor}}))
=\dim(\Sigma_{\Psi})
=
\begin{cases}
  1& r_{0}=0\\
  2& r_{0}>0
\end{cases}\ .
$$

There are several extensions from the Lundquist form.
One was studied in \cite{Vandas2003} with the elliptic cylindrical coordinates,
$$
x=c\cosh u\cos v,\qquad
y=c\sinh u\sin v,\qquad
z=z,
$$
where $u\geq0$, $v\in[0,2\pi)$, and $c$ is constant.  
The induced Riemannian metric in this coordinate system is
$$
\bm{g}=c^{2}(\cosh^{2}u-\cos^{2}v)(\bdr u\otimes\bdr u+\bdr v\otimes\bdr v)
+\bdr z\otimes\bdr z.
$$
The contact form was constructed with special functions.

\subsubsection{The standard contact form on $S^{3}$}

To consider a contact form on
$$
S^{3}=\{ (x^{1},y^{1},x^{2},y^{2})\in\mbbR^{4}\ |\
(x^{1})^{2}+(y^{1})^{2}+(x^{2})^{2}+(y^{2})^{2}=1 \},
$$
firstly introduce 
$$
\bm{\lambda}^{\prime}
=x^{1}\bdr y^{1}-y^{1}\bdr x^{1}+x^{2}\bdr y^{2}-y^{2}\bdr x^{2},
$$
on $\mbbR^{4}$ with coordinates $(x^{1},y^{1},x^{2},y^{2})$.
Let $\iota:S^{3}\to\mbbR^{4}$ be the embedding, and let 
$$
\bm{\lambda}=\iota^{*}\bm{\lambda}^{\prime}.
$$
Then $(S^{3},\ker\bm{\lambda})$ is a contact manifold. The Reeb vector field is
$$
R=x^{1}\frac{\partial}{\partial y^{1}}
-y^{1}\frac{\partial}{\partial x^{1}}
+x^{2}\frac{\partial}{\partial y^{2}}
-y^{2}\frac{\partial}{\partial x^{2}}.
$$
A non-trivial first integrals of $R$ is found as
$$
\Psi(x^{1},y^{1},x^{2},y^{2})
=\Psi_{1}(x^{1},y^{1}),\qquad
\Psi_{1}(x^{1},y^{1})
:=(x^{1})^{2}+(y^{1})^{2}.
$$
Note that although
$\Psi_{2}:=(x^{2})^{2}+(y^{2})^{2}$ is also a first integral,  
$\Psi_{1}$ and $\Psi_{2}$ are not independent due to $\Psi_{2}+\Psi_{1}=1$. 
Therefore only $\Psi$ is considered below. 

To explicitly express 
the critical point set and its Hessian in coordinates,
introduce the so-called toroidal coordinates: 
\begin{align}
x^{1}&=\sin\theta_{3}\cos\theta_{1},
\non\\
y^{1}&=\sin\theta_{3}\sin\theta_{1},
\non\\
x^{2}&=\cos\theta_{3}\cos\theta_{2},
\non\\
y^{2}&=\cos\theta_{3}\sin\theta_{2},
\non  
\end{align}
where
$$
0\leq\theta_{1}<2\pi,\qquad
0\leq\theta_{2}<2\pi,\qquad
0\leq\theta_{3}\leq\frac{\pi}{2}.
$$
The Riemannian metric induced from the standard metric in $\mbbR^{4}$ is
\begin{align}
\bm{g}
&=\sin^{2}\theta_{3}\bdr\theta_{1}\otimes\bdr\theta_{1}
+\cos^{2}\theta_{3}\bdr\theta_{2}\otimes\bdr\theta_{2}
+\bdr\theta_{3}\otimes\bdr\theta_{3}
\non\\
&=\bm{\sigma}^{1}\otimes\bm{\sigma}^{1}
+\bm{\sigma}^{2}\otimes\bm{\sigma}^{2}
+\bm{\sigma}^{3}\otimes\bm{\sigma}^{3},
\non
\end{align}
where $\{\bm{\sigma}^{1},\bm{\sigma}^{2},\bm{\sigma}^{3}\}$ is a coframe.  
This coframe and its dual $\{e_{1},e_{2},e_{3}\}$ are 
\begin{align}
\bm{\sigma}^{1}
&:=\sin\theta_{3}\bdr\theta_{1},&
e_{1}
&=\frac{1}{\sin\theta_{3}}\frac{\partial}{\partial\theta_{1}},
\non\\
\bm{\sigma}^{2}
&:=\cos\theta_{3}\bdr\theta_{2},&
e_{2}
&=\frac{1}{\cos\theta_{3}}\frac{\partial}{\partial\theta_{2}},
\non\\
\bm{\sigma}^{3}
&:=\bdr\theta_{3},&
e_{3}
&=\frac{\partial}{\partial\theta_{3}}.
\non
\end{align}
The contact form and its Reeb vector field are written as
\begin{align}
\bm{\lambda}
&=\sin^{2}\theta_{3}\bdr\theta_{1}
+\cos^{2}\theta_{3}\bdr\theta_{2},
\non\\
&=\sin\theta_{3}\bm{\sigma}^{1}+\cos\theta_{3}\bm{\sigma}^{2},
\non\\
R&=\frac{\partial}{\partial\theta_{1}}+\frac{\partial}{\partial\theta_{2}},
\non\\
&=\sin\theta_{3}e_{1}+\cos\theta_{3}e_{2}.
\non
\end{align}
The non-trivial first integral of $R$, $\Psi=\Psi_{1}=(x^{1})^{2}+(x^{2})^{2}$,
is written as 
$$
\Psi
=\sin^{2}\theta_{3},
$$
and then the critical point set is found
from $\bdr\Psi=2\sin\theta_{3}\cos\theta_{3}\bdr \theta_{3}$ as 
$$
\Sigma_{\Psi}
=\Sigma_{\Psi,0}\cup
\Sigma_{\Psi,\pi/2},
$$
where
\begin{align}
\Sigma_{\Psi,0}
&:=\{\ (\theta_{1},\theta_{2},\theta_{3})\ |\ \theta_{3}=0 \ \}
=\{\ (0,0,x^{2},y^{2})\ |\ (x^{2})^{2}+(y^{2})^{2}=1 \ \},
\non\\
\Sigma_{\Psi,\pi/2}
&:=\left\{\ (\theta_{1},\theta_{2},\theta_{3})\ \bigg| \
\theta_{3}=\frac{\pi}{2} \ \right\}
=\{\ (x^{1},y^{1},0,0)\ |\ (x^{1})^{2}+(y^{1})^{2}=1 \ \}.
\non
\end{align}
The Hessian is evaluated at a point of $\Sigma_{\Psi,0}$ and that of
$\Sigma_{\Psi,\pi/2}$ as 
$$
\text{Hess}_{p}\Psi
=\begin{cases}
\ \ 2\bdr\theta_{3}\otimes\bdr\theta_{3}&\text{for}\quad p\in \Sigma_{\Psi,0}
\\
-2\bdr\theta_{3}\otimes\bdr\theta_{3}&\text{for}\quad  p \in\Sigma_{\Psi,\pi/2}
\end{cases} .
$$
Hence $\Psi$ is a Morse-Bott function: 
$$
T_{p}\Sigma_{\Psi}
=\ker(\text{Hess}_{p}\Psi)
=\text{span}\left\{\frac{\partial}{\partial \theta_{1}},
\frac{\partial}{\partial\theta_{2}}
\right\},\qquad
\forall p\in\Sigma_{\Psi}. 
$$
To obtain the horizontal contact Hamiltonian vector field, 
decompose $W_{\text{hor}}$ 
in terms of the basis of $\ker\bm{\lambda}$ as
\begin{align}
W_{\text{hor}}
&=
a_{12}\left(
\cos^{2}\theta_{3}\frac{\partial}{\partial\theta_{1}}
-\sin^{2}\theta_{3}\frac{\partial}{\partial\theta_{2}}
\right)
+a_{3}\frac{\partial}{\partial \theta_{3}}
\non\\
&=a_{12}\sin\theta_{3}\cos\theta_{3}
(\cos\theta_{3}e_{1}-\sin\theta_{3}e_{2})+a_{3}e_{3},
\non
\end{align}
with $a_{12}$ and $a_{3}$ being some functions to be determined.
Straightforward calculations show that
\begin{align}
\bdr\bm{\lambda}
&=2\sin\theta_{3}\cos\theta_{3}
(\bdr\theta_{3}\wedge\bdr\theta_{1}+\bdr\theta_{2}\wedge\bdr\theta_{3}),
\non\\
&=2\cos\theta_{3}\bm{\sigma}^{3}\wedge\bm{\sigma}^{1}
+2\sin\theta_{3}\bm{\sigma}^{2}\wedge\bm{\sigma}^{3},
\non\\
\sstar\bdr\bm{\lambda}
&=2\cos\theta_{3}\bm{\sigma}^{2}+2\sin\theta_{3}\bm{\sigma}^{1},
\non\\
&=2\bm{\lambda}.
\non
\end{align}
In addition, it follows that
\begin{align}
\ii_{W_{\text{hor}}}\bdr\bm{\lambda}
&=2\cos\theta_{3}\ii_{W_{\text{hor}}}(\bm{\sigma}^{3}\wedge\bm{\sigma}^{1})
+2\sin\theta_{3}\ii_{W_{\text{hor}}}(\bm{\sigma}^{2}\wedge\bm{\sigma}^{3})
\non\\
&=2\cos\theta_{3}(a_{3}\bm{\sigma}^{1}-a_{12}\cos^{2}\theta_{3}\sin\theta_{3}
\bm{\sigma}^{3})
-2\sin\theta_{3}(a_{12}\sin^{2}\theta_{3}\cos\theta_{3}\bm{\sigma}^{3}+a_{3}\bm{\sigma}^{2})
\non\\
&=2a_{3}(\cos\theta_{3}\bm{\sigma}^{1}-\sin\theta_{3}\bm{\sigma}^{2})
-2a_{12}\sin\theta_{3}\cos\theta_{3}\bm{\sigma}^{3}
\non\\
\bdr\Psi
&=2\sin\theta_{3}\cos\theta_{3}\bm{\sigma}^{3}.
\non
\end{align}
These show that $a_{3}=0$ and $a_{12}=1$,
such that $\ii_{W_{\text{hr}}}\bdr\bm{\lambda}=-\bdr\Psi$: 
$$
W_{\text{hor}}
=\sin\theta_{3}\cos\theta_{3}
(\cos\theta_{3}e_{1}-\sin\theta_{3}e_{2}).
$$
Straightforward calculations show that   
$$
\bm{g}(W_{\text{hor}},W_{\text{hor}})
=\sin^{2}\theta_{3}\cos^{2}\theta_{3},
$$
which yields that $W_{\text{hor}}|_{p}=0$ at $p\in\Sigma_{\Psi}$.  
In addition,  from straightforward calculations 
it follows that $R$ and $W_{\text{hor}}$ are orthogonal:
\begin{align}
  \bm{g}(R,-)
  &=\sin\theta_{3}\bm{\sigma}^{1}+\cos\theta_{3}\bm{\sigma}^{2},
  \non\\
  \bm{g}(R,W_{\text{hor}})
  &=0.
\non
\end{align}
Since $\Psi$ is a Morse-Bott function, it follows that
$$
\dim(\text{Fix}(W_{\text{hor}}))
=\dim(\Sigma_{\Psi})
=2.
$$ 

\subsubsection{The ABC flow}

Let $N=T^{3}=S^{1}\times S^{1}\times S^{1}$
with the coordinates $(x^{1},x^{2},x^{3})=(x,y,z)$,
$0\leq x,y,z<2\pi$, and
$$
\bm{g}
=\bdr x\otimes\bdr x+\bdr y\otimes \bdr y+\bdr z\otimes \bdr z.
$$
Consider the one-form 
\beq
\bm{\lambda}
=(A\sin z+C\cos y)\bdr x
+(B\sin x+A\cos z)\bdr y
+(C\sin y+B\cos x)\bdr z,
\label{ABC-general}
\eeq
where $A,B$ and $C$ are constant. 
This one-form is rotational Beltrami with $\kappa=1$, 
$$
\sstar\bdr\bm{\lambda}
=\bm{\lambda}.
$$
The Reeb vector field is
$$
R=\frac{A\sin z+C\cos y}{\bm{g}^{-1}(\bm{\lambda},\bm{\lambda})}
\frac{\partial}{\partial x}
+\frac{B\sin x+A\cos z}{\bm{g}^{-1}(\bm{\lambda},\bm{\lambda})}
\frac{\partial}{\partial y}
+\frac{C\sin y+B\cos x}{\bm{g}^{-1}(\bm{\lambda},\bm{\lambda})}
\frac{\partial}{\partial z},
$$
which shows chaotic trajectories in general values of $A,B$ and $C$.
Flow associated with \fr{ABC-general}
is called the {\it Arnold–Beltrami–Childress (ABC) flow}.  
See \cite{Maciejewski2002} for non-integrability on this model.  
Although there is no proof, a non-trivial first integral of $R$
is not expected to be found in this general case.
So we focus on a special case. 

When $C=0$, the contact form and Reeb vector reduce to 
\begin{align}
  \bm{\lambda}
  &=(A\sin z)\bdr x
+(B\sin x+A\cos z)\bdr y
+(B\cos x)\bdr z,
\non\\
R&=\frac{A\sin z}{\bm{g}^{-1}(\bm{\lambda},\bm{\lambda})}
\frac{\partial}{\partial x}
+\frac{B\sin x+A\cos z}{\bm{g}^{-1}(\bm{\lambda},\bm{\lambda})}
\frac{\partial}{\partial y}
+\frac{B\cos x}{\bm{g}^{-1}(\bm{\lambda},\bm{\lambda})}
\frac{\partial}{\partial z},
\non
\end{align}
where $A$ and $B$ are chosen so that 
\begin{align}
\bm{g}^{-1}(\bm{\lambda},\bm{\lambda})
&=A^{2}\sin^{2}z+B^{2}\sin^{2}x+A^{2}\cos^{2}z+2AB\sin x\cos z+B^{2}\cos^{2}x
\non\\
&=A^{2}+B^{2}+2AB\sin x\cos z
\non\\
&\neq 0,
\non
\end{align}
for all $x$ and $z$. In this case, one observes that 
$$
\cL_{\partial/\partial y}\bm{\lambda}
=(B\sin x+A\cos z)\cL_{\partial/\partial y}\bdr y
=0.
$$
This and Proposition\,\ref{fact-how-to-find-first-integral} yield that 
$$
\Psi(x,z)
=\ii_{\partial/\partial y}\bm{\lambda}
=B\sin x+A\cos z,
$$
is a non-trivial first integral of $R$.
This function is Morse-Bott, which is verified from 
\begin{align}
  \bdr\Psi
  &=B\cos x\bdr x-A\sin z\bdr z,
  \non\\
  \Sigma_{\Psi}
  &=\{(x,y,z)|x=(m_{x}+1/2)\pi,z=m_{z}\pi,\ m_{x},m_{z}\in \mbbZ\},
\non\\
\text{Hess}_{p}\Psi
&=-B\sin x\bdr x\otimes \bdr x-A\cos z\bdr z\otimes\bdr z,
\qquad \forall p\in\Sigma_{\Psi},
\non
\end{align}
as
$$
T_{p}\Sigma_{\Psi}
=\ker(\text{Hess}_{p}\Psi)
=\text{span}\left\{\frac{\partial}{\partial y}\right\},
\qquad \forall p\in\Sigma_{\Psi}.
$$
To obtain the horizontal contact Hamiltonian vector field $W_{\text{hor}}$,
decompose it in terms of the basis of $\ker\bm{\lambda}$. The vector space $\ker\bm{\lambda}$ is spanned by  
$$
B\cos x\frac{\partial}{\partial x}-A\sin z\frac{\partial }{\partial z},\qquad\text{and}\qquad
\frac{\partial}{\partial y}-\Psi R,
$$
and then the decomposition is of the form 
\begin{align}
W_{\text{hor}}
&=a_{1}\left(B\cos x\frac{\partial}{\partial x}-A\sin z\frac{\partial}{\partial z}\right)
+a_{2}\left(\frac{\partial}{\partial y}-\Psi R\right), 
\non
\end{align}
with $a_{1}$ and $a_{2}$ being some functions to be determined.
Then it turns out that $a_{1}=0$ and $a_{2}=1$, so that
$\ii_{W_{\text{hor}}}\bdr\bm{\lambda}=-\bdr\Psi$: 
\begin{align}
\ii_{\partial/\partial y-\Psi R}\bdr\bm{\lambda}
&=\ii_{\partial/\partial y}\bdr\bm{\lambda}-\Psi\ii_{R}\bdr\bm{\lambda}
\non\\
&=\ii_{\partial/\partial y}\bdr\bm{\lambda}
\non\\
&=\ii_{\partial/\partial y}\left[
A\sin z\bdr y\wedge\bdr z
  +\Psi\bdr z\wedge\bdr x
  +B\cos x\bdr x\wedge\bdr y
\right]
\non\\
&=A\sin z\bdr z
-B\cos x\bdr x
\non\\
&=-\bdr\Psi.
\non
\end{align}
The explicit form of $W_{\text{hor}}$ is thus 
$$
W_{\text{hor}}
=\frac{\partial}{\partial y}
-\frac{\Psi(x,z)}{\bm{g}^{-1}(\bm{\lambda},\bm{\lambda})}\left[
A\sin z\frac{\partial}{\partial x}
+\Psi(x,z)\frac{\partial}{\partial y}
+B\cos x\frac{\partial}{\partial z}\right].
$$
With $\bm{g}$,  
one obtains the orthogonal relation between $R$ and $W_{\text{hor}}$:
$$
\bm{g}(R,W_{\text{hor}})
=0,
$$
which is verified as 
\begin{align}
\bm{g}(-,W_{\text{hor}})
&=\left(1-\frac{\Psi^{2}}{\bm{g}^{-1}(\bm{\lambda},\bm{\lambda})}\right)\bdr y
-\frac{\Psi}{\bm{g}^{-1}(\bm{\lambda},\bm{\lambda})}
\left(A\sin z\bdr x+B\cos x\bdr z\right),
\non\\
\bm{g}(R,W_{\text{hor}})
&=\left(1-\frac{\Psi^{2}}{\bm{g}^{-1}(\bm{\lambda},\bm{\lambda})}\right)(Ry)
-\frac{\Psi}{\bm{g}^{-1}(\bm{\lambda},\bm{\lambda})}
\left[A\sin z(Rx)+B\cos x(Rz)\right]
\non\\
&=\frac{\Psi}{\bm{g}^{-1}(\bm{\lambda},\bm{\lambda})}
\left[1-\frac{\Psi^{2}+A^{2}\sin^{2} z+B^{2}\cos^{2} x}{\bm{g}^{-1}(\bm{\lambda},\bm{\lambda})}\right]
\non\\
&=\frac{\Psi}{\bm{g}^{-1}(\bm{\lambda},\bm{\lambda})}
\left[1-\frac{A^{2}+B^{2}+2AB\sin x\cos z}{\bm{g}^{-1}(\bm{\lambda},\bm{\lambda})}\right]
\non\\
&=0.
\non
\end{align}
Since $\Psi$ is a Morse-Bott function, it follows that
$$
\dim(\text{Fix}(W_{\text{hor}}))
=\dim(\Sigma_{\Psi})
=1.
$$ 

\section{Discussions and conclusions}
\label{section-dicussion-conclusions}

Throughout this paper, symplectic and contact geometries have been employed
to describe electromagnetic fields, and an advantage of this approach is
as follows. From a purely mathematical interest,
three-dimensional contact manifolds and four-dimensional symplectic manifolds
have been extensively studied. These studies have led to
the discovery of numerous theorems\,\cite{McDuff,Hofer1994,Silva2008}, and
such theorems ought to be more widely applied in physics.
Based on this background, this paper has demonstrated 
the applicability of modern geometry to the long-standing study
of electromagnetic fields, finding a new solution of Maxwell's equations and its applications.
Indeed, the use of the 3+1 decomposition of Maxwell
fields corresponds to a symplectization that is highly standard
in contact and symplectic geometries.
This correspondence has shown to be useful,
which is one piece of evidence for 
this approach. Note that in this approach,   
studying electromagnetic fields at a fixed time is equivalent to
studying a class of vector fields on a contact manifold. Therefore, 
applications of recent developments of contact geometry
to electromagnetism should be beneficial.

As stated in Theorem\,\ref{fact-introduction-1},
a class of Maxwell fields with sources has explicitly been constructed,   
where the sources and Maxwell fields are generated by 
non-singular rotational Beltrami fields.
This theorem leads to how to amplify electromagnetic energy density
by tuning the external source, and it leads to a clue for clarifying
how to generate magnetic fields observed in the solar wind\,\cite{Vandas2017}.
In addition, as stated in Corollary\,\ref{fact-closed-field-lines}, 
by applying the resolution of the Weinstein conjecture\,\cite{Hutchings2010}
that is a well-known result in contact geometry, the existence of closed field
lines has been shown when the space manifold $N$ is closed. 
Since closed field lines play a role of backbones of field configurations,
this statement on the existence is a step toward
understanding complicated field lines. 

As stated in Theorem\,\ref{fact-introduction-2},
a class of solutions to Maxwell's equations and relations with superconductors
have been shown.
Since superconductors have many engineering applications,
further theoretical developments are always required. 
A solution to the London equation
given in Corollary\,\ref{fact-London-equations} is one of such directions. 
To advance our understanding of this solution from a viewpoint of microscopic physics, a detailed study of a microscopic toy model and the solution
is required. In addition, a comparison with a solution
in the context of the Einstein equation\,\cite{Flores2026}
is one key to further understand the present solution. 

As stated in Theorem\,\ref{fact-introduction-3}, 
the existence of the so-called magnetic stream function for fields 
in Theorems\,\ref{fact-introduction-1} and \ref{fact-introduction-2} implies
that the Maxwell system can be viewed as a Liouville integrable system, where
the Liouville integrability is originated from Hamiltonian mechanics.  
Since there are many mathematical statements on
integrable Hamiltonian systems, it is expected that these existing theorems
help to understand magnetic surfaces. 
Since magnetic surfaces are considered when plasma 
is studied, this perspective helps to advance research in plasma science.

The developments in symplectic and contact geometries 
are expected to be applied to 
various sciences and engineering in general. 
On the one hand, such applications were found in 
the study of
thermodynamics\,\cite{Mrugala1990,Goto2015,Bravetti2017,Schaft2018,Entov2026}, 
the ideal fluid\,\cite{Etnyre2000,Cardona2019},
general relativity\,\cite{Kozaki2022,Kozaki2024}, and 
three-body problems\,\cite{Moreno2022,Frauenfelder2018}.
On the other hand, other applications in electromagnetism 
are less known\,\cite{Goto2024,Dahl2008}. Therefore, various fundamental
theorems are expected to be found by pursuing research in this direction.  
There are several potential future works that follow from this paper. 
They include the following:
\begin{itemize}
\item
  generalizing the present work of Maxwell systems 
  to systems 
  with non-trivial media\,\cite{Dahl2013,Gratus2020}, 

\item
  finding applications of the present study in electromagnetism and
  plasma physics,  

\item
  applying known theorems in the study of Bott integrability\,\cite{Geiges2024}
  to electromagnetism, 
  
\item
  finding more relations between the Maxwell fields in this paper and
  fields expressed in the original London equations. 

\end{itemize}

We believe that the elucidation of these remaining questions together
with this work 
will develop the geometric theories 
of electromagnetism and their applications to various related sciences.

\subsection*{Data availability statement}

No new data were created or analyzed in this study.

\subsection*{Acknowledgment}
The author would like to thank Yosuke Nakata and David A Burton 
for pointing out unclear points in an earlier version of the paper
and for offering suggestions to improve it.
A discussion regarding the London equations 
was added to the paper after Y Nakata provided his comments.

\appendix 
\section{Appendix}
\label{section-appendix}
In this appendix, 
details of derivations of equations in the main text, 
and examples of rotational Beltrami fields are shown. 

\subsection{Decomposed form of Maxwell's equations}
\label{section-appendix-Maxwell-decomposed}
In the following the decomposed form of Maxwell's equations
\fr{Maxwell-decomposed-vacuum} is derived. 

Substituting \fr{F0} into 
\fr{Maxwell-equations-vacuum}, one has that  
\beqa
\dr F
&=&- \dr(c_{0}\bm{B}+\bm{e}\wedge\dr x^{0})
=- c_{0}\left(\dr t\wedge\cL_{\partial/\partial t}\bm{B}
+\bdr\bm{B}\right)
-\bdr\bm{e}\wedge\dr x^{0}
\non\\
&=&-\left(\cL_{\partial/\partial t}\bm{B}+\bdr\bm{e}
\right)\wedge\dr x^{0}-c_{0}\bdr\bm{B}
=0.
\non
\eeqa
Note that $\bdr\bm{B}$ does not contain $\dr x^{0}$. 
Decomposing the above equation, $\dr F=0$,
into the term proportional to $\dr x^{0}$ and the other term, 
one obtains the first and second equations of \fr{Maxwell-decomposed-vacuum}.

Substituting \fr{F0} into 
\fr{Maxwell-equations-vacuum} with $\bm{B}=\mu_{0}\sstar\bm{h}$ by
\fr{Maxwell-constitutive-relations-vacuum-decomposed} and
Lemma\,\ref{fact-star-decompositions-1},
one has that
\begin{align}
\star F
&=-c_{0}\mu_{0}\star\sstar\bm{h}+\star (\dr x^{0}\wedge\bm{e})
\non\\
&=c_{0}\mu_{0}\bm{h}\wedge\dr x^{0}-\sstar\bm{e},
\non
\end{align}
from which
$$
\varepsilon_{0}\star F
=c_{0}^{-1}\bm{h}\wedge\dr x^{0}-\bm{D}.
$$
Hence, 
\begin{align}
\varepsilon_{0}\dr\star F
&=-\dr(\bm{D}-c_{0}^{-1}\bm{h}\wedge\dr x^{0})
\non\\
&=-\dr t\wedge\cL_{\partial/\partial t}\bm{D}-\bdr\bm{D}
+c_{0}^{-1}\bdr\bm{h}\wedge\dr x^{0}
\non\\
&=-c_{0}^{-1}\left(\dot{\bm{D}}-\bdr\bm{h}
\right)\wedge\dr x^{0}-\bdr\bm{D}.
\non
\end{align}
Note that $\bdr\bm{D}$ does not contain $\dr x^{0}$. 
Decomposing 
the above equation, $\varepsilon_{0}\dr\star F=j_{4}$, with \fr{j0}:
$$
j_{4}=-\rho\sstar 1+c_{0}^{-1}\bm{J}\wedge\dr x^{0},
$$
into the term propositional to $\dr x^{0}$ and the other term, 
one obtains the last two equations of \fr{Maxwell-decomposed-vacuum}.

\subsection{Derivation of \fr{e-h-from-F}}
\label{section-derivation-e-h-from-F}

In the following the equations in \fr{e-h-from-F} are derived.

Recalling that $F$ is given by \fr{F0} and
$\ii_{\partial/\partial x^{0}}\bm{B}=\ii_{\partial/\partial x^{0}}\bm{e}=0$,
one has that
\begin{align}
\ii_{\partial/\partial x^{0}}F
&=\ii_{\partial/\partial x^{0}}(-c_{0}\bm{B}+\dr x^{0}\wedge\bm{e})
\non\\
&=\bm{e}.
\non
\end{align}

Recalling that $F$ is given by \fr{F0}, 
$\bm{B}=\mu_{0}\sstar\bm{h}$ by
\fr{Maxwell-constitutive-relations-vacuum-decomposed},  
using $\wt{\partial/\partial x^{0}}=-\dr x^{0}$, and
Lemma\,\ref{fact-star-decompositions-1},  
one has that
\begin{align}
\ii_{\partial/\partial x^{0}}\star F
&=-\star(F\wedge\dr x^{0})
\non\\
&=c_{0}\star(\bm{B}\wedge\dr x^{0})
\non\\
&=c_{0}\mu_{0}\star(\sstar\bm{h}\wedge\dr x^{0})
\non\\
&=-c_{0}\mu_{0}\ii_{\partial/\partial x^{0}}\star(\sstar\bm{h})    
\non\\
&=-c_{0}\mu_{0}\ii_{\partial/\partial x^{0}}(\dr x^{0}\wedge\bm{h})
\non\\ 
&=-c_{0}\mu_{0}\bm{h}.
\non 
\end{align}
Combining the obtained expression and
$c_{0}\mu_{0}= \sqrt{\mu_{0}/\varepsilon_{0}}=\varepsilon_{0}^{-1}c_{0}^{-1}$,
one arrives at
$$
\bm{h}=-\sqrt{\frac{\varepsilon_{0}}{\mu_{0}}}\ii_{\partial/\partial x^{0}}\star F
=-\varepsilon_{0}c_{0}\ \ii_{\partial/\partial x^{0}}\star F.
$$

\subsection{Several explicit solutions with Theorem\,\ref{fact-introduction-1}}
\label{section-fact-introduction-1-several}

Observe that the solution to \fr{ode1} is decomposed as
$f=f_{\hom}+f_{\inhom}$,  where $f_{\hom}$ is the so-called
homogeneous solution, $\ddot{f}_{\hom}+c_{0}^{2}\kappa^{2}f_{\hom}=0$, and
$f_{\inhom}$ is the so-called inhomogeneous solution. 

A few classes of solutions $F$ can be constructed by tuning $f_{\ext}$ in
Theorem\,\ref{fact-introduction-1}.
\begin{enumerate}
\item 
Choose $f_{\ext}(t)=f_{\ext}(0)\sin(c_{0}\kappa t)$. 
Since the frequency of $f_{\ext}(t)$ matches with the natural frequency
of \fr{ode1}, a solution $f$ to \fr{ode1}
in Theorem\,\ref{fact-introduction-1} has    
resonance behavior.
Hence the consequence of Theorem\,\ref{fact-introduction-1} is that 
the amplitudes of electromagnetic fields
can be amplified by a thorough choice of $f_{\ext}$ with
the resonance property of \fr{ode1}. 
With this property, one has a system in which 
electromagnetic energy increases in time
(see Corollary\,\ref{fact-consequence-1} below). 
\item
Choose $f_{\ext}(t)=\nu \kappa^{2}t$ with $\nu\in\mbbR$ being constant,
so that
$j_{4}=\varepsilon_{0}(\sstar\bm{\lambda}_{\kappa})\wedge \nu\kappa^{2} c_{0}^{-1} x^{0}\dr x^{0}$.
Observe that the component of $j_{4}$ is proportional to $x^{0}$,
and it physically expresses that the source increases linearly in $x^{0}$.   
In this case, it follows that
$$
f(t)=-\nu t,
$$
is a solution to \fr{ode1}. Correspondingly, the two-form
$$
F=-\nu c_{0}^{-1}\dr(x^{0}\bm{\lambda}_{\kappa}),
$$
is a solution to Maxwell's equations.
When the homogeneous solution of \fr{ode1} is absent, one finds that
$$
\bm{e}=-\nu c_{0}^{-1}\bm{\lambda}_{\kappa},\qquad
\bm{D}=-\nu \varepsilon_{0}c_{0}^{-1}\sstar\bm{\lambda}_{\kappa},\qquad
\bm{h}=\varepsilon_{0}\kappa\nu x^{0}\bm{\lambda}_{\kappa},\qquad
\bm{B}=c_{0}^{-2}\kappa\nu x^{0}\sstar\bm{\lambda}_{\kappa},
$$
(see \fr{Maxwell-constitutive-relations-vacuum-decomposed} and \fr{e-h-from-F}
below for the derivations).  
Notice that the amplitudes of $\bm{h}$ and $\bm{B}$ increase linearly in 
$x^{0}$.  
With this property, one has a system in which 
electromagnetic energy increases in time
(see Corollary\,\ref{fact-consequence-2} below). 

\item
Choose $f_{\ext}(t)=f_{\ext}(0)=\nu$, for all $t$, so that
$j_{4}=\varepsilon_{0}(\sstar\bm{\lambda}_{\kappa})\wedge\nu\dr x^{0}$.  
Observe that the component of $j_{4}$ is independent of $x^{0}$. 
In this case, it follows that
$$
f(t)=f(0)=-\nu\kappa^{-2}=\const,
$$
is a solution to \fr{ode1}. Correspondingly, the two-form 
$$
F=-\nu\kappa^{-2}\bdr\bm{\lambda}_{\kappa}, 
$$
is a time independent solution to Maxwell's equations,
in the sense of $\cL_{\partial/\partial t}F=0$.
When the homogeneous solution of \fr{ode1} is absent,   
since $\ii_{\partial/\partial x^{0}}F=0$,
one finds the trivial $\bm{e}$ and $\bm{D}$, together  
with the non-trivial $\bm{h}$ and $\bm{B}$:  
$$
\bm{e}=0,\qquad
\bm{D}=0,\qquad 
\bm{h}=\varepsilon_{0}c_{0}\nu\kappa^{-1}\bm{\lambda}_{\kappa},\qquad
\bm{B}=c_{0}^{-1}\nu\kappa^{-1}\sstar\bm{\lambda}_{\kappa},
$$
where  \fr{Maxwell-constitutive-relations-vacuum-decomposed} and
\fr{e-h-from-F} have been used for the derivations.  
\end{enumerate}

\subsection{Application of Theorem\,\ref{fact-introduction-1}: Amplification of  fields}

There are several ways to control the electromagnetic fields in general,
and one is to design the source $j_{4}$ in Maxwell's equations. 
Here, it is briefly argued how $f_{\ext}$ in Theorem\,\ref{fact-introduction-1}
affects the electromagnetic fields. In particular,  
how to amplify the energy densities is discussed. 
To this end the solution of \fr{ode1},
$$
f(t)=f_{0}\cos(\omega_{\kappa}t)+f_{1}\sin(\omega_{\kappa}t)
-\frac{c_{0}^{2}}{\omega_{\kappa}}\int_{t_{0}}^{t}f_{\ext}(t^{\prime})
\sin(\omega_{\kappa}(t-t^{\prime}))\dr t^{\prime},
$$
is employed where $t_{0},f_{0},f_{1}$ are constant, and 
$$
\omega_{\kappa}
:=c_{0}\kappa. 
$$
Since $\kappa\neq0$ and $c_{0}>0$, one has that $\omega_{\kappa}\neq 0$.
In what follows $t_{0}=0$ is kept fixed. 

Various behavior of the electromagnetic fields and that of their total energy
densities can be realized by choosing $f_{\ext}$.
This is summarized in Table\,\ref{table-type-input}.

\begin{table}
\caption{Types of the input and their properties of the total energies}
\label{table-type-input}
\begin{center}
\begin{tabular}{|l|l|l|}
\hline
Type of the input $f_{\ext}(t)$ & Solution $f (t)$ & Total energy  \\ \hline\hline
$0$
& \fr{ode1-sin-input-homogeneous-solution}
& Conserved \\ \hline 
$f_{\ext}(0)\sin(\omega_{\ext}t),\quad 
(\omega_{\ext}\neq \omega_{\kappa}, \omega_{\ext}>0, f_{\ext}(0)\neq 0)$  
& \fr{ode1-sin-input-general-solution-non-resonance}
& Sinusoidal  \\ \hline
$f_{\ext}(0)\sin(\omega_{\ext}t),\quad 
(\omega_{\ext}=\omega_{\kappa},  f_{\ext}(0)\neq 0)$  
& \fr{ode1-sin-input-general-solution-resonance}
& Amplified  \\ \hline
$\nu\kappa^{2}t,\quad
(\nu=\const)$  
& \fr{ode1-t-input-general-solution}
& Amplified  \\ \hline
$\nu,\quad (\nu=\const)$  
& \fr{ode1-constant-input-general-solution}
& Sinusoidal   \\ \hline
\end{tabular}
\end{center}
\end{table}

\subsubsection{$f_{\ext}(t)=f_{\ext}(0)\sin(\omega_{\ext} t)$ with $\omega_{\ext}>0$ constant}

Choose
$$
f_{\ext}(t)
=f_{\ext}(0)\sin(\omega_{\ext} t),
$$
with $f_{\ext}(0)$ and $\omega_{\ext}>0$ being constant.
Then \fr{ode1} is written as
\beq
\ddot{f}+\omega_{\kappa}^{2}f
=-c_{0}^{2}f_{\ext}(0)\sin(\omega_{\ext}t).
\label{ode1-sin-input}
\eeq
Then several properties can be shown as follows. 
\begin{itemize}
\item
(No input case, $f_{\ext}(0)=0$). 
Consider the case of $f_{\ext}(0)=0$, that corresponds to $j_{4}=0$. 
The electromagnetic fields are explicitly obtained by solving the ODE,  
$$
\ddot{f}_{\hom}+\omega_{\kappa}^{2}f_{\hom}
=0. 
$$
This ODE is of homogeneous type, and its solution is obtained as  
\beq
f_{\hom}(t)
=A_{0}\cos(\omega_{\kappa}t)
+A_{1}\sin(\omega_{\kappa}t),
\label{ode1-sin-input-homogeneous-solution}
\eeq
where $A_{0}:=f_{\hom}(0)$ and $A_{1}:=\dot{f}_{\hom}(0)/\omega_{\kappa}$
are constant. 
Straightforward
calculations with \fr{ode1-sin-input-homogeneous-solution} yield
$$
\left(c_{0}^{-1}\dot{f}_{\hom}\right)^{2}+(\kappa f_{\hom})^{2}
=\kappa^{2}\left(A_{0}^{2}+A_{1}^{2}\right).
$$
Then the total energy density three-form
reduces from \fr{total-energy-3-form-theorem1} to 
\begin{align}
\bm{\cE}_{\emem}^{f_{\hom}}
&:=\frac{\varepsilon_{0}}{2}\left\{
(c_{0}^{-1}\dot{f}_{\hom})^{2}+(\kappa f_{\hom})^{2}
\right\}\bm{\lambda}_{\kappa}\wedge\sstar\bm{\lambda}_{\kappa}
\non\\
&=\frac{\varepsilon_{0}}{2}\kappa^{2}\left(A_{0}^{2}+A_{1}^{2}\right)
\bm{\lambda}_{\kappa}\wedge\sstar\bm{\lambda}_{\kappa},
\non
\end{align}
which is time independent.  
Therefore the integral of $\bm{\cE}_{\emem}^{f_{\hom}}$ over a non-empty 
region  $N^{\prime}\subset N$ is conserved in time: 
$$
\frac{\dr}{\dr t}\int_{N^{\prime}}  \bm{\cE}_{\emem}^{f_{\hom}}
=0,\qquad \forall t.
$$

\item
(Non-resonance case, $\omega_{\ext}\neq \omega_{\kappa}$).\ 
The solution to \fr{ode1-sin-input} with $\omega_{\ext}\neq \omega_{\kappa}$
is given by
\beq
f(t)
=f_{\hom}(t)+A_{2}\sin(\omega_{\ext}t), \quad
A_{2}
:=-\,\frac{c_{0}^{2}}{\omega_{\kappa}^{2}-\omega_{\ext}^{2}}
f_{\ext}(0),
\label{ode1-sin-input-general-solution-non-resonance}
\eeq
where $f_{\hom}(t)$ has been given by
\fr{ode1-sin-input-homogeneous-solution}. 
Simple calculations with \fr{ode1-sin-input-general-solution-non-resonance}
yield 
\begin{align}
\left(c_{0}^{-1}\dot{f}\right)^{2}+(\kappa f)^{2}
&=  \left(c_{0}^{-1}\dot{f}_{\hom}\right)^{2}+(\kappa f_{\hom})^{2}
\non\\
&+2A_{2}\left(
c_{0}^{-2}\omega_{\ext}\dot{f}_{\hom}\cos(\omega_{\ext}t)
+\kappa^{2}f_{\hom}\sin(\omega_{\ext}t)
\right)
\non\\
&+A_{2}^{2}\left(
c_{0}^{-2}\omega_{\ext}^{2}\cos^{2}(\omega_{\ext}t)
+\kappa^{2}\sin^{2}(\omega_{\ext}t)\right),
\non
\end{align}
and
\begin{align}
\frac{\dr}{\dr t}\left[\left(c_{0}^{-1}\dot{f}\right)^{2}+(\kappa f)^{2}
  \right]
&=2A_{2}\frac{\dr}{\dr t}\left(
c_{0}^{-2}\omega_{\ext}\dot{f}_{\hom}\cos(\omega_{\ext}t)
+\kappa^{2}f_{\hom}\sin(\omega_{\ext}t)
\right)
\non\\
&+A_{2}^{2}\frac{\dr}{\dr t}\left(
c_{0}^{-2}\omega_{\ext}^{2}\cos^{2}(\omega_{\ext}t)
+\kappa^{2}\sin^{2}(\omega_{\ext}t)\right),
\non\\
&=2A_{2}c_{0}^{-2}\omega_{\ext}\left(
-\omega_{\kappa}^{2}f_{\hom}\cos(\omega_{\ext}t)
-\omega_{\ext}\dot{f}_{\hom}\sin(\omega_{\ext}t)
\right)
\non\\
&+2A_{2}\kappa^{2}\left(\dot{f}_{\hom}\sin(\omega_{\ext}t)+\omega_{\ext}f_{\hom}\cos(\omega_{\ext}t)\right)
\non\\
&+2A_{2}^{2}\left(
-c_{0}^{-2}\omega_{\ext}^{2}+\kappa^{2}\right)
\cos(\omega_{\ext}t)\sin(\omega_{\ext}t).
\non
\end{align}
Therefore the integral of $\bm{\cE}_{\emem}^{f}$,
\fr{total-energy-3-form-theorem1}, over a non-empty
region $N^{\prime}\subset N$ yields 
$$
\frac{\dr}{\dr t}\int_{N^{\prime}}\bm{\cE}_{\emem}^{f}
\neq 0.
$$
The right hand side of the above equation 
show sinusoidal modulation as a function of time.
  
\item
(Resonance case, $\omega_{\ext}=\omega_{\kappa}$).\ 
The solution to \fr{ode1-sin-input} with $\omega_{\ext}=\omega_{\kappa}$
is given by
\beq
f(t)
=f_{\hom}(t)+f_{\inhom}(t),\qquad
f_{\inhom}(t)
:=\frac{c_{0}^{2}}{2\omega_{\kappa}}
f_{\ext}(0) t\,\cos(\omega_{\kappa}t),
\label{ode1-sin-input-general-solution-resonance}
\eeq
where $f_{\hom}(t)$ has been given by 
\fr{ode1-sin-input-homogeneous-solution}. 

The total energy density 
three-form reduces from \fr{total-energy-3-form-theorem1} to 
\begin{align}
\bm{\cE}_{\emem}^{f}
&=\bm{\cE}_{\emem}^{f_{\hom}}
+\frac{\varepsilon_{0}}{2}\frac{c_{0}f_{\ext}(0)}{\omega_{\kappa}}
\left[
c_{0}^{-2}\dot{f}_{\hom}(\cos(\omega_{\kappa}t)-t\omega_{\kappa}\sin(\omega_{\kappa}t))+\kappa^{2}f_{\hom}t\cos(\omega_{\kappa}t)
\right]
\non\\
&+  \frac{\varepsilon_{0}}{2}
\left(\frac{c_{0}^{2}f_{\ext}(0)}{2\omega_{\kappa}}\right)^{2}
\left( \kappa^{2}t^{2}+c_{0}^{-2}
\cos^{2}(\omega_{\kappa}t)-2c_{0}^{-2}\omega_{\kappa}t\cos(\omega_{\kappa}t)
\sin(\omega_{\kappa}t)
\right)
\bm{\lambda}_{\kappa}\wedge\sstar\bm{\lambda}_{\kappa}.
\label{total-energy-resonance-case1}
\end{align}
\end{itemize}

As a consequence of Theorem\,\ref{fact-introduction-1}, 
energy in a non-empty region $N^{\prime}\subset N$ is amplified. 
\begin{Corollary}[Energy amplification]  
\label{fact-consequence-1}
In the resonance case above, 
the integral of $\bm{\cE}_{\emem}^{f}$ over a non-empty 
region $N^{\prime}\subset N$ is proportional to $t^{2}$ for $t\gg 1$:
$$
\int_{N^{\prime}}  \bm{\cE}_{\emem}^{f}
\propto t^{2},\qquad t\gg1,
$$
and hence the total energy increases in time:  
$$
\frac{\dr}{\dr t}\int_{N^{\prime}}  \bm{\cE}_{\emem}^{f}
\propto t,\qquad t\gg1.
$$
\end{Corollary}
\begin{Proof}
It is immediately obtained from 
\fr{total-energy-resonance-case1}.
\qed
\end{Proof}

\subsubsection{$f_{\ext}(t)=\nu\kappa^{2}t$ with $\nu$ constant}
Choose 
$$
f_{\ext}(t)
=\nu\kappa^{2}t,
$$
with $\nu$ being constant.
In this case, 
\beq
f(t)=f_{\hom}(t)-\nu t, 
\label{ode1-t-input-general-solution}
\eeq
is a solution to \fr{ode1}. This yields the following.
\begin{Corollary}[Energy amplification]
\label{fact-consequence-2}
In the case of $f_{\ext}(t)=\nu\kappa^{2}t$,
the integral of $\bm{\cE}_{\emem}^{f}$ over a non-empty 
region $N^{\prime}\subset N$ is proportional to $t^{2}$ for $t\gg 1$:
$$
\int_{N^{\prime}}  \bm{\cE}_{\emem}^{f}
=\int_{N^{\prime}}  \bm{\cE}_{\emem}^{f_{\hom}}
+\left\{-\varepsilon_{0}\nu
\left(c_{0}^{-2}\dot{f}_{\hom}+\kappa^{2} t f_{\hom}\right)
+\frac{\varepsilon_{0}\nu^{2}}{2}\frac{1+\omega_{\kappa}^{2}t^{2}}{c_{0}^{2}}
\right\}\int_{N^{\prime}}
\bm{\lambda}_{\kappa}\wedge\sstar\bm{\lambda}_{\kappa},
$$
and hence the integral of $\cE_{\emem}^{f}$ over $N^{\prime}$ is not conserved:
$$
\frac{\dr}{\dr t}\int_{N^{\prime}}  \bm{\cE}_{\emem}^{f}
=\left[
  -\varepsilon_{0}\nu\kappa^{2}t \dot{f}_{\hom}
+\frac{\varepsilon_{0}\nu^{2}}{c_{0}^{2}}\omega_{\kappa}^{2}t
\right]
\int_{N^{\prime}}
\bm{\lambda}_{\kappa}\wedge\sstar\bm{\lambda}_{\kappa}.
$$
When $f_{\hom}$ is absent, it follows that 
$$
\frac{\dr}{\dr t}\int_{N^{\prime}}  \bm{\cE}_{\emem}^{f}
\propto t,\qquad
t\gg1. 
$$
\end{Corollary}
\begin{Proof}
It is obtained from 
straightforward calculations with \fr{total-energy-3-form-theorem1}.
\qed
\end{Proof}

\subsubsection{$f_{\ext}(t)=\nu$ with $\nu$ constant}
Choose
$$
f_{\ext}(t)
=\nu,
$$
with $\nu$ being constant.
In this case, 
\beq
f(t)=f_{\hom}(t)-\kappa^{-2}\nu,  
\label{ode1-constant-input-general-solution}
\eeq
is a solution to \fr{ode1}.
This with \fr{total-energy-3-form-theorem1} yield  
\begin{align}
\bm{\cE}_{\emem}^{f}
&=\bm{\cE}_{\emem}^{f_{\hom}}
+\frac{\varepsilon_{0}}{2}
\left(
\kappa^{-2}\nu^{2}
-2\nu f_{\hom}\right)
\bm{\lambda}_{\kappa}\wedge\sstar\bm{\lambda}_{\kappa},
\non
\end{align}
which is time independent.
Therefore the integral of $\bm{\cE}_{\emem}^{f}$ over a non-empty 
region  $N^{\prime}\subset N$ yields 
$$
\frac{\dr}{\dr t}\int_{N^{\prime}}  \bm{\cE}_{\emem}^{f}
\neq 0,\qquad \forall t.
$$
The integral of $\bm{\cE}_{\emem}^{f}$ over $N^{\prime}$
and its time derivative show
sinusoidal modulation as a function of time.

\begin{Remark}
When $f_{\hom}(0)=0$, 
$\bm{e}$ and $\bm{D}$ are trivial (or vanishing), whereas
$\bm{h}$ and $\bm{B}$ are non-trivial:
$$
\bm{e}=0,\qquad
\bm{D}=0,\qquad 
\bm{h}=\varepsilon_{0}c_{0}\kappa^{-1}\bm{\lambda}_{\kappa},\qquad
\bm{B}=c_{0}^{-1}\kappa^{-1}\sstar\bm{\lambda}_{\kappa},
$$
which have been derived from \fr{e-from-f}, \fr{h-from-f}, and
\fr{Maxwell-constitutive-relations-vacuum-decomposed}. 
This means that a static magnetic field with vanishing electric field
is realized in this case.  
\end{Remark}

\subsection{Examples of Beltrami one-form}
\label{section-example-Beltrami-form}

The following are examples of rotational Beltrami fields.
\begin{Example}
\label{example-Beltrami-Giroux}
Consider the three-torus $N=T^{3}=S^{1}\times S^{1}\times S^{1}$, the
metric on $T^{3}$ as the induced metric from  \fr{g-E^3}, 
$$
\bm{g}
=\bdr x^{1}\otimes\bdr x^{1}
+\bdr x^{2}\otimes\bdr x^{2}
+\bdr x^{3}\otimes\bdr x^{3}.
$$
and $\sstar1=\bdr x^{1}\wedge\bdr x^{2}\wedge \bdr x^{3}$. 
Let $\bm{\lambda}_{-n}$ be the one-form with a fixed $n\in\mbbZ_{>0}$ 
$$
\bm{\lambda}_{-n}
=c(\,\cos(nx^{3})\bdr x^{1}+\sin(nx^{3})\bdr x^{2}),
$$
with $c>0$ being constant.  
See Theorem 4.1 of \cite{Etnyre2000} for a significance
of this $\bm{\lambda}_{-n}$. 
This $\bm{\lambda}_{-n}$ has the property that 
$$
\bm{g}^{-1}(\bm{\lambda}_{-n},\bm{\lambda}_{-n})
=c^{2}.
$$
Then the one-form $\bm{\lambda}_{-n}$ satisfies
$$
\sstar\bdr\bm{\lambda}_{-n}
=-n\bm{\lambda}_{-n}. 
$$
\end{Example}
\begin{Example}
Consider the three-torus $N=T^{3}$. Let $\bm{g}$ and $\sstar 1$ be given 
as the ones in Example\,\ref{example-Beltrami-Giroux}. 
Let $\bm{\lambda}_{\kappa}$ be the one-form such that
$$
\bm{\lambda}_{\kappa}
=\left(A\sin (\kappa x^{3})+C\cos (\kappa x^{2})\right)\bdr x^{1}
+\left(B\sin (\kappa x^{1})+A\cos (\kappa x^{3})\right)\bdr x^{2}
+\left(C\sin (\kappa x^{2})+B\cos (\kappa x^{1})\right)\bdr x^{3},  
$$
with $A,B$, and $C$ being constant. The metric dual of $\bm{\lambda}_{1}$, 
$Z_{1}=\wt{\bm{\lambda}_{1}}^{\bm{g}}$, is called 
the {\it Arnold–Beltrami–Childress (ABC) flow} 
(see \cite{Dombre1986} for applications with $\kappa=1$ in fluid mechanics).  
Straightforward calculations show that 
$$
\sstar\bdr\bm{\lambda}_{\kappa}=\kappa\bm{\lambda}_{\kappa}.
$$
\end{Example}
\begin{Example}
Consider the solid torus $D^{2}\times S^{1}$. 
The coordinates for $D^{2}$ are set to be $r$ and $\varphi$ so that 
$(x^{1},x^{2})=(r\cos\varphi,r\sin\varphi)$ with
$r^{2}=(x^{1})^{2}+(x^{2})^{2}$,
and that for $S^{1}$ is set to be $x^{3}$. 
The metric is chosen to be the induced one from \fr{g-E^3}
$$
\bm{g}
=\bdr r\otimes\bdr r+r^{2}\bdr \varphi\otimes \bdr\varphi
+\bdr x^{3}\otimes\bdr x^{3},
$$
and let $\sstar1=r\bdr r\wedge\bdr \varphi\wedge\bdr x^{3}$.  
The actions of the Hodge map are as follows:
$$
\sstar\dr r
=r\bdr\varphi\wedge\bdr x^{3},\quad
\sstar(r\dr\varphi)
=\bdr x^{3}\wedge\bdr r,\quad
\sstar\bdr x^{3}
=\bdr r\wedge r\bdr\varphi,
$$
and
$$
\sstar(r\bdr \varphi\wedge\bdr x^{3})
=\bdr r,\quad
\sstar(\bdr x^{3}\wedge\bdr r)
=r\bdr \varphi,\quad
\sstar(\bdr r\wedge r\bdr\varphi)
=\bdr x^{3}.
$$
Let $\bm{\lambda}_{-k}$ be the one-form
$$
\bm{\lambda}_{-k}
=\frac{\beta}{k_{c}}J_{1}(k_{c}r)\sin(\beta x^{3})\bdr r
-\frac{k}{k_{c}}J_{1}(k_{c}r) \cos(\beta x^{3})r \bdr \varphi
+J_{0}(k_{c}r)\cos(\beta x^{3})\bdr x^{3},
$$
where $J_{n}(z)$ is the $n$-th Bessel function of the first kind,
$k_{c}>0$ and $\beta$ are constant, 
and 
$k^{2}=\beta^{2}+k_{c}^{2}$. 
There are several formulae for the Bessel functions. 
Using these formulae, one has that 
$$
\sstar\bdr \bm{\lambda}_{-k}
=- k \bm{\lambda}_{-k}.
$$
See \cite{Mochizuki2022} for applications of these fields in electromagnetism. 
\end{Example}


\end{document}